\documentclass[a4paper,11pt]{article}
\usepackage{jheppub}
\usepackage{amssymb}
\usepackage{amsmath}
\usepackage{graphicx}
\usepackage{multirow}
\usepackage{subfigure}
\usepackage{float}
\usepackage{graphics}
\usepackage{slashed}
\usepackage{booktabs}

\allowdisplaybreaks
\def\be{\begin{equation}}
\def\ee{\end{equation}}
\newcommand{\bea}{\begin{equation} \begin{array}{c}}
\newcommand{\eea}{ \end{array} \end{equation}}

\def\as{\alpha_s}

\def\e{\epsilon}

\def\nno{\nonumber}

\def\hlinew#1{%
  \noalign{\ifnum0=`}\fi\hrule \@height #1 \futurelet
   \reserved@a\@xhline}
\title{Threshold resummation effects in Higgs boson pair production at the LHC}

\author[a]{Ding Yu Shao,}
\author[a,b]{Chong Sheng Li,}
\author[a]{Hai Tao Li}
\author[a]{and Jian Wang}

\affiliation[a]{School of Physics and State Key Laboratory of
Nuclear Physics and Technology, Peking University,\\Beijing 100871,
China}
\affiliation[b]{Center for High Energy Physics,
Peking University,\\Beijing 100871, China}

\emailAdd{csli@pku.edu.cn}

\abstract{We investigate the resummation effects in the Standard Model Higgs boson pair production through gluon-gluon fusion at the LHC with soft-collinear effective theory. We calculate the total cross section and the invariant mass distribution at Next-to-Next-to-Leading-Logarithmic level with $\pi^2$-enhanced terms resummed, which are matched to the QCD Next-to-Leading Order results. In the high order QCD predictions exact top quark mass effects are included in full form factors. Our results show that the resummation effects increase the Next-to-Leading Order results by about $20\% \sim 30\%$, and the scale uncertainty is reduced to $8\%$, which lead to increased confidence on the theoretical predictions. The PDF+$\as$ uncertainties are almost not changed after including resummation effects. We also study the sensitivities of the total cross section and the invariant mass distribution to the Higgs boson self-coupling. We find that the total cross section and the invariant mass distribution shape depend strongly on the Higgs boson self-coupling, and therefore it is possible to extract Higgs boson self-coupling from the total cross section and invariant mass distribution when the measurement precision increases at the LHC.}

\begin{document}
\maketitle
\flushbottom

\section{INTRODUCTION}\label{s1}
Recently, at the CERN Large Hadron Collider~(LHC), both the ATLAS\cite{:2012gk} and CMS\cite{:2012gu} collaborations have found an about $5\sigma$ excess for a Standard Model (SM) Higgs boson like particle with a mass around $125~{\rm GeV}$. In the future, it is promising that LHC can tell whether this particle is the SM Higgs boson or something else when the measurement precision increases and its properties are probed.

In the SM, the Higgs boson is responsible for the origin of electroweak symmetry breaking and the generation of elementary particle masses. After the Higgs field $\Phi$ gets the vacuum expectation value $v$, the SM Higgs potential in the unitary gauge can be written as
\begin{eqnarray}
 V(h)= \lambda \left[\frac{(v+h)^2}{2}-\frac{v^2}{2}\right]^2,
\end{eqnarray}
where the Higgs boson self-coupling $\lambda$ is given by $\lambda_{\rm SM}=m_H^2/(2v^2)$ at the tree-level in the SM, and the radiative corrections can decrease $\lambda_{\rm SM}$ by $10\%$ for $m_{H}=125$ GeV where main contributions from top quark loops~\cite{Kanemura:2002vm}. In some new physics models, if a new heavy particle has the similar non-decoupling property to the top quark, much larger radiative corrections may exit. For example, in Two-Higgs-Doublet-Model, due to the non-decoupling effects of the additional heavier Higgs boson in loops, one-loop corrections can cause the lightest Higgs boson self-coupling to deviate from SM prediction by about $100\%$, even if the lightest Higgs boson couplings with gauge bosons and fermions, respectively, are almost SM-like~\cite{Kanemura:2002vm}. Thus, it is important to measure the value of Higgs boson self-coupling to distinguish the SM from other models.

At the LHC, the Higgs boson self-coupling $\lambda$ can be directly probed through Higgs boson pair production via gluon gluon fusion, and the relevant studies have been performed~\cite{Glover:1987nx,Plehn:1996wb,Dawson:1998py,Djouadi:1999rca,Baglio:2012np,Binoth:2006ym,Baur:2002rb,Baur:2002qd,Baur:2003gpa,Baur:2003gp,Dolan:2012rv,Papaefstathiou:2012qe}. In Ref.~\cite{Dawson:1998py}, the Next-to-Leading Order~(NLO) QCD corrections have been calculated in the large top quark mass limit, and the K-factor is found to be large~(about 2), but the scale uncertainty at the NLO is still very high~(about 30\%). Hence, it is important to investigate higher order effects and perform QCD resummation calculations in order to improve the predictions on the cross section and reduce the theoretical uncertainties. Moreover, as is shown in Ref.\cite{Baur:2003gp}, the Higgs boson pair invariant mass distribution strongly depends on the Higgs boson self-coupling $\lambda$. Thus in order to improve the precision of theoretical predictions it is also necessary to calculate QCD resummation effects on the invariant mass distribution. Besides, compared to single Higgs boson production, due to the larger invariant mass in the final state, gluons radiated from initial states are much softer. Thus, the soft gluon resummation effects are expected to be much more important than in the case of single Higgs boson production.

The full NLO QCD calculations for Higgs boson pair production including exact top quark mass effects are absent at present. The approximated method of NLO QCD calculations applied in previous works is multiplying the Leading-Order~(LO) cross section for a finite top quark mass by a K-factor obtained in the large top quark mass limit~\cite{Baur:2002rb,Baur:2002qd,Baur:2003gpa,Baur:2003gp,Dolan:2012rv,Papaefstathiou:2012qe}. This approximated method works well in the single Higgs boson production~\cite{Spira:1995rr}. We will work in the infinite top quark mass limit, but the top quark mass effects are partly included in the exact top quark mass dependent form factors. The same approach has been also applied in the publicly available code \texttt{HPAIR}~\cite{hpair}.

In this work, we perform the threshold resummation in Higgs boson pair production at the LHC using soft-collinear effective theory~(SCET)~\cite{Bauer:2000ew,Nikolaev:2000sh,Bauer:2001ct,Bauer:2001yt,Becher:2006nr}, which is very suitable to deal with the scattering processes with multiple scales. In the time-like process the threshold region is usually defined as the limit $z=M^2/s\to1$, where $M$ is the invariant mass of the time-like particle and $s$ is the square of the partonic center-of-mass energy\cite{Idilbi:2005ky,Becher:2007ty,Ahrens:2008qu,Ahrens:2008nc,Mantry:2009qz,Zhu:2009sg,Idilbi:2009cc,Yang:2006gs,Zhu:2010mr}. Generally, in the threshold region, the cross section can be factorized as
 \begin{eqnarray}
  \sigma = \mathcal{H} \otimes \mathcal{S} \otimes f_{P_a} \otimes f_{P_b},
 \end{eqnarray}
where $\mathcal{H}$, $\mathcal{S}$ and $f_{P}$ are the hard function, soft function and parton distribution function~(PDF), respectively. The hard function incorporates the short distance contributions arising from virtual corrections. The soft gluon effects coming from all colored particles are contained in the soft function. The PDF denotes the probability of finding a particular parton in the proton. In the SCET, the soft and collinear degrees of freedom can decouple via field redefinition, so the physics at different scales can be studied separately~\cite{Bauer:2001yt}. One can deal with the relevant degrees of freedom at each scale and their scale dependences are controlled by renormalization group~(RG) equations, respectively. As a result, the large logarithmic terms between different scales can be resummed conveniently.

\indent The arrangement of this paper is as follows. In Sec.\ref{s2}, we give the fixed order~(FO) results of Higgs boson pair production at the LHC. In Sec.\ref{s3}, we calculate the hard and soft matching coefficients at the NLO, and present RG-improved differential cross section analytically.
In Sec.\ref{s4}, we discuss the choice of hard and soft matching scale, and the numerical results of total cross section and the invariant mass distribution of Higgs boson pair. We conclude in Sec.\ref{s5}.

\section{FO RESULTS}\label{s2}

\begin{figure}[h]
 \begin{center}
  \includegraphics[width=1\textwidth]{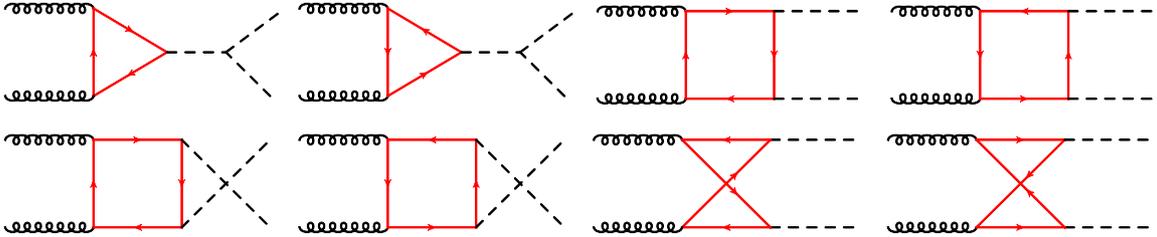}\\
  \caption{Feynman diagrams for Higgs boson pair production induced by top quark loop at the Born level.}\label{Fig:fin_mt_born}
  \end{center}
\end{figure}

We consider the process of Higgs boson pair production at the LHC
\begin{eqnarray}
 P(P_1) + P(P_2) \rightarrow h^0(p_3) + h^0(p_4) + X(P_X),
\end{eqnarray}
where $X$ is the final hadronic state. In the Born approximation the Higgs boson pairs are mainly produced through gluon-gluon fusion mediated by the top quark loop,
\begin{eqnarray}
 g(p_1) + g(p_2) \rightarrow h^0(p_3) + h^0(p_4),
\end{eqnarray}
where $p_1 = x_1 P_1$ and $p_2 = x_2 P_2$. The corresponding Feynman diagrams are shown in Fig.~\ref{Fig:fin_mt_born}. We define the kinematic invariants
\begin{eqnarray}
 && S = (P_1 + P_2)^2, ~~~ s = (p_1 + p_2)^2, ~~~ M^2 = (p_3 + p_4)^2, \nno \\
 && t = (p_1 - p_3)^2, ~~~ u = (p_2 - p_3)^2.
\end{eqnarray}

At the Born level, the triple differential cross section for the variables the rapidity $Y$, the invariant mass $M$ of Higgs boson pair and the angle $\theta$ between $\vec{p}_1$ and $\vec{p}_3$ in the partonic center-of-mass frame can be written as
\begin{eqnarray}\label{fin_mt_born}
 \frac{d^3\sigma}{dM^2 dY d\cos\theta} &=& \frac{\as^2 G_{\rm F}^2 M^2 \beta_H}{2304 \left(2\pi\right)^3 S} f_{g/P}(x_1, \mu_f) f_{g/P}(x_2, \mu_f)
 \Bigg[ \Big|f_{\rm Tri}^{\rm A}(M^2, t, u, m_t^2, m_H^2)  \nno \\
 && + f_{\rm Box}^{\rm A}(M^2, t, u, m_t^2, m_H^2)\Big|^2 + \Big| f_{\rm Box}^{\rm B}(M^2, t, u, m_t^2, m_H^2) \Big|^2 \Bigg],
\end{eqnarray}
where $f_{g/P}$ is the gluon Parton Distribution Function~(PDF) for initial proton $P$ and $\mu_f$ is the factorization scale. The 3-velocity $\beta_H$ of $h^0$ in the final state Higgs boson pair rest frame is defined as
\begin{eqnarray}
 \beta_H = \sqrt{1 - \frac{4 m_H^2}{M^2} }.
\end{eqnarray}
We calculated the triangular and box form factor $f_{\rm Tri}^{\rm A}$, $f_{\rm Box}^{\rm A}$ and $f_{\rm Box}^{\rm B}$and found agreement with Refs.~\cite{Glover:1987nx,Plehn:1996wb}. Their expressions are collected in App.~\ref{a1}. In the infinite top quark mass limit $m_t \rightarrow +\infty$, these form factors read
\begin{align}\label{formfac_lmt}
 f_{\rm Tri}^{\rm A} \rightarrow \frac{6 \lambda v^2}{ M^2 - m_H^2 + i m_H \Gamma_H }, ~~~ f_{\rm Box}^{\rm A} \rightarrow -1, ~~~  f_{\rm Box}^{\rm B} \rightarrow 0,
\end{align}
where $\Gamma_H$ is the total decay width of Higgs boson.

\begin{figure}
 \begin{center}
  \includegraphics[width=0.4\textwidth]{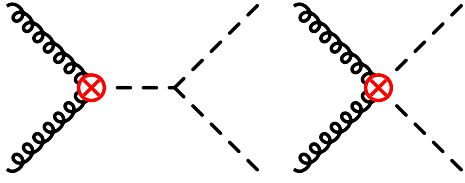}\\
  \caption{Feynman diagrams for Higgs boson pair production at the tree level in the infinite top quark mass limit.}\label{born}
  \end{center}
\end{figure}

The full NLO QCD calculations for Higgs boson pair production including exact top quark mass effects are absent at present. In Ref.~\cite{Dawson:1998py} the NLO QCD corrections to the total cross section are investigated in the infinite top quark mass limit. However, the NLO differential distributions are not available there. In order to obtain more precise kinematic information of Higgs boson pair, we calculate the triple differential distribution at the NLO QCD level. In the calculations we partly include exact top quark mass effects at the LO by using the triangular and box form factor $f_{\rm Tri}^{\rm A}$, $f_{\rm Box}^{\rm A}$ and $f_{\rm Box}^{\rm B}$. The same approach has been also applied in the publicly available code \texttt{HPAIR}~\cite{hpair}.

Firstly, we study the Higgs boson pair production in the infinite top quark mass limit. The relevant effective Lagrangian describing $ggh$ and $gghh$ interactions~\cite{Dawson:1998py} can be written as
\begin{eqnarray}\label{effL}
 \mathcal{L}_{\rm eff}=\frac{\as(\mu^2)}{12\pi v}C_t(\mu^2)G_{\mu\nu}^a G^{a~\mu\nu}h - \frac{\as(\mu^2)}{24\pi v^2}C_t(\mu^2)G_{\mu\nu}^a G^{a~\mu\nu}h^2
\end{eqnarray}
where the Higgs vacuum expectation value $v$ is related to the Fermi constant by $v=\left(\sqrt{2}G_F\right)^{-1/2}$. At the LO, Feynman diagrams for Higgs boson pair production induced by above effective Lagrangian are shown in Fig.~\ref{born}. Up to NLO, the Wilson coefficient $C_t(\mu^2)$ are derived by performing the large top quark mass expansion of the the corresponding one- and two-loop Feynman diagrams in Ref.~\cite{Dawson:1998py}, and can be written as
\begin{eqnarray}\label{Ct}
 C_t(\mu^2)=1+\frac{11}{4}\frac{\as(\mu^2)}{\pi}.
\end{eqnarray}
In the derivations of the Wilson coefficient $C_t(\mu^2)$, the scale hierarchy is assumed
\begin{eqnarray}\label{hierachy}
 m_t^2 \gg s, M^2, |t|, |u|, m_H^2.
\end{eqnarray}
Going beyond the Born level, the triple differential cross section can be written as
\begin{eqnarray}
 \frac{d^3\sigma}{dM^2dYd\cos\theta} &=& \frac{\as^2 G_{\rm F}^2 M^2 \beta_H}{2304(2\pi)^3S} \sum_{i j} \int_{0}^{1}dy\int_{\bar{z}(y)}^{1}\frac{dz}{z}
 f_{i/A}(x_1,\mu_f)f_{j/B}(x_2,\mu_f) \nno \\
 &&\times C_{ij}(z,y,M,m_t,m_H,\cos\theta,\mu_f).
\end{eqnarray}
The hard scattering kernel $C_{ij}(z,y,M,m_t,m_H,\cos\theta,\mu_f)$ depends on variables $z$ and $y$, which are defined as
\begin{eqnarray}
 z=\frac{M^2}{s},~~~~y=\frac{\frac{x_1}{x_2}e^{-2Y}-z}{(1-z)(1+\frac{x_1}{x_2}e^{-2Y})}.
\end{eqnarray}

\begin{figure}[H]
 \begin{center}
  \includegraphics[width=0.5\textwidth]{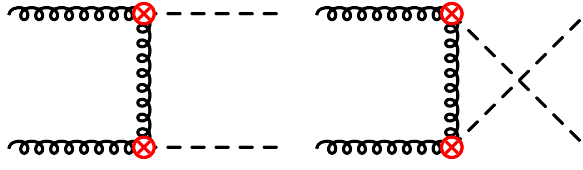}\\
  \caption{A part of virtual corrections for Higgs boson pair production at the NLO in the infinite top quark mass limit, which have neither ultraviolet nor infrared divergences.}\label{vt2}
  \end{center}
\end{figure}

At the LO, the combination $(ij)=(gg)$, and at the NLO, $(ij)=(gg),(qg),(\bar{q}g),(gq),(g\bar{q}),(q\bar{q}),(\bar{q}q)$. The parton momentum fractions $x_1$ and $x_2$ can be written as
\begin{eqnarray}
 x_1=\sqrt{\frac{\tau}{z}\frac{1-(1-y)(1-z)}{1-y(1-z)}}e^Y,~~~~
 x_2=\sqrt{\frac{\tau}{z}\frac{1-y(1-z)}{1-(1-y)(1-z)}}e^{-Y}.
\end{eqnarray}
In order to make sure that they are not exceed 1, the lower limit of integration of the variable $z$ is defined as
\begin{eqnarray}
 \bar{z}(y)=\max\Bigg\{&&\frac{(\tau e^{2Y}-1)(1-y)+\sqrt{4\tau y^2e^{2Y}+(1-y)^2(1-\tau e^{2Y})^2}}{2y}, \nno\\
 && \frac{(\tau e^{-2Y}-1)y + \sqrt{4\tau(1-y)^2e^{-2Y} + y^2(1-\tau e^{-2Y})^2}}{2(1-y)} \Bigg\},
\end{eqnarray}
where the variable $\tau$ is defined as $\tau=M^2/S$.
\begin{figure}
 \begin{center}
  \includegraphics[width=0.7\textwidth]{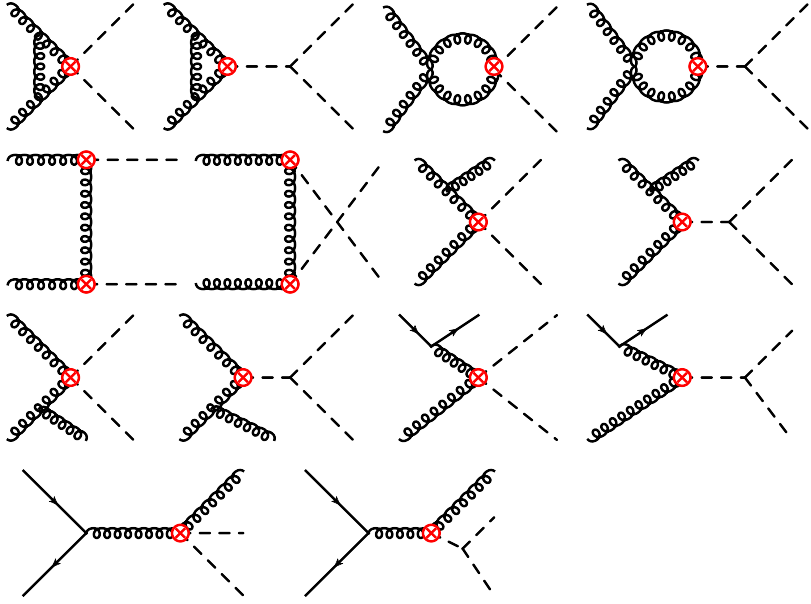}\\
  \caption{Feynman diagrams for Higgs boson pair production at the NLO in the infinite top quark mass limit.}\label{feyndia}
  \end{center}
\end{figure}
Up to the NLO, the explicit results of $C_{ij}(z,y,M,m_t,m_H,\cos\theta,\mu_f)$ can be written as follows, for which the corresponding Feynman diagrams are shown in Fig.~\ref{feyndia}.
\begin{eqnarray}\label{cij_def}
 C_{gg} &=& \delta(1-z)\frac{\delta(y)+\delta(1-y)}{2}
 \Bigg\{\left[1+\frac{\as}{\pi}\left(2\pi^2 + \frac{11}{2}\right)\right]
 \Bigg[ \Big|f_{\rm Tri}^{\rm A} + f_{\rm Box}^{\rm A}\Big|^2 + \Big| f_{\rm Box}^{\rm B} \Big|^2 \Bigg] \nno\\
% && - \frac{\as}{\pi}\frac{2}{3}\mathcal{R}e\left(1-\frac{6\lambda v^2}{M^2-m_H^2+i \Gamma_H m_H}\right)\Bigg\}
 && + \frac{\as}{\pi} \mathcal{R}e\Bigg[ \left( f_{\rm Tri}^{\rm A} + f_{\rm Box}^{\rm A} \right) g^{\rm A\ast} + f_{\rm Box}^{\rm B} g^{\rm B \ast} \Bigg]
 + \frac{\as}{\pi}\Bigg\{ \frac{\delta(y)+\delta(1-y)}{2}
 \Bigg[ 6D(z) \nno\\
 &&+ 6\ln\frac{M^2(1-z)^2}{\mu_f^2z}\left(\frac{1}{z}-2+z-z^2\right) \Bigg]
 + 3\left[\frac{1}{1-z}\right]_+\left(\left[\frac{1}{y}\right]_+ + \left[\frac{1}{1-y}\right]_+ \right) \nno\\
 &&\times \frac{(1-z)^4(y^4-2y^3+3y^2-2y)+(1-z+z^2)^2}{z}
% + 3\left(\left[\frac{1}{y}\right]_+ + \left[\frac{1}{1-y}\right]_+ \right)\frac{(1-z+z^2)^2}{z(1-z)} \nno\\
% &&- \frac{(1-z)^3(y^2-y+2)}{z}
  \Bigg\} \Bigg[ \Big|f_{\rm Tri}^{\rm A} + f_{\rm Box}^{\rm A}\Big|^2 + \Big| f_{\rm Box}^{\rm B} \Big|^2 \Bigg], \nno \\
 \\
 C_{qg} &=& \frac{\as}{\pi}\Bigg\{\frac{2}{3}\delta(1-y)\left[\frac{1+(1-z)^2}{z}
 \ln\frac{M^2(1-z)^2}{\mu_f^2 z} + z \right] +
 \frac{2}{3}\frac{1+(1-z)^2}{z}\left[\frac{1}{1-y}\right]_+ \nno\\
 && - \frac{2}{3}\frac{(1+y)(1-z)^2}{z} \Bigg\}
 \Bigg[ \Big|f_{\rm Tri}^{\rm A} + f_{\rm Box}^{\rm A}\Big|^2 + \Big| f_{\rm Box}^{\rm B} \Big|^2 \Bigg], \\
 C_{q\bar{q}} &=& \frac{\as}{\pi}\frac{16(1-z)^3}{9z}(1-2y+2y^2)
 \Bigg[ \Big|f_{\rm Tri}^{\rm A} + f_{\rm Box}^{\rm A}\Big|^2 + \Big| f_{\rm Box}^{\rm B} \Big|^2 \Bigg],
\end{eqnarray}
where we set the renormalization scale to be the factorization scale $\mu_f$. Here $D(z)$ denotes the plus distribution as follows,
\begin{eqnarray}\label{Dz}
 D(z)=\left[\frac{1}{1-z}\ln\frac{M^2(1-z)^2}{\mu_f^2z}\right]_+.
\end{eqnarray}
The form factors $g^A$ and $g^B$ describe parts of virtual corrections, which have neither ultraviolet nor infrared divergences. The corresponding Feynman diagrams are shown in Fig.~\ref{vt2}. The expressions of $g^A$ and $g^B$ containing complete top quark mass dependence are listed in App.~\ref{a1}. In the infinite top quark mass limit, we have
\begin{eqnarray}
 g^A & \longrightarrow & \frac{2}{3}, \\
 g^B & \longrightarrow & - \frac{p_T^2}{3 t u}(M^2 - 2 m_H^2),
\end{eqnarray}
where $p_T^2 = (u t - m_H^4)/s$. In Ref.~\cite{Dawson:1998py} the approximated form factors $g^A$ and $g^B$ in the infinite top quark mass limit are used. Here in order to retain more kinematic information we will use the complete expressions of form factors. We have checked that the numerical differences for the total cross section in these two cases are small and can be neglected.

\noindent The other $C_{ij}$ can be obtained from the symmetry relations
\begin{eqnarray}
 C_{\bar{q}g}=C_{qg},~~~C_{gq}=C_{g\bar{q}}=C_{qg}|_{y\to1-y},~~~C_{\bar{q}q}=C_{q\bar{q}}.
\end{eqnarray}
We can split the integral kernel $C_{gg}$ as a sum of singular terms and regularized terms~\cite{Becher:2007ty},
\begin{eqnarray}
 C_{gg} &=& \frac{\delta(y)+\delta(1-y)}{2} \Bigg[ \Big|f_{\rm Tri}^{\rm A} + f_{\rm Box}^{\rm A}\Big|^2 + \Big| f_{\rm Box}^{\rm B} \Big|^2 \Bigg] C(z,M,\mu_f) + C_{gg}^{reg},
\end{eqnarray}
where the singular terms consist of the contributions from the virtual and real soft gluons, and have the explicit form
\begin{align}\label{Ckernal}
  C(z,M,\mu_f) &= \delta(1-z) + \frac{\as}{\pi}\left[ \left(2\pi^2 + \frac{11}{2}\right)\delta(1-z) + 6D(z) \right].
%  C(z,y,M,m_t,m_H,\cos\theta,\mu_f) &= \delta(1-z) + \frac{\as}{\pi}\left[ \left(2\pi^2 + \frac{11}{2}\right)\delta(1-z) + 6D(z) \right] \nno \\
%  &\hspace{-2em}  + \frac{\as}{\pi} \delta(1-z) \mathcal{R}e \left[ \frac{  \left( f_{\rm Tri}^{\rm A} + f_{\rm Box}^{\rm A} \right) g^{\rm A\ast} + f_{\rm Box}^{\rm B} g^{\rm B \ast} }{  \left|f_{\rm Tri}^{\rm A} + f_{\rm Box}^{\rm A}\right|^2 + \left| f_{\rm Box}^{\rm B} \right|^2  } \right]
\end{align}
After performing the integration over $y$, we can get the singular differential cross section in the threshold region
\begin{align}
\label{cs}
\frac{d^3\sigma^{\rm Singular}}{dM^2dYd\cos\theta} & = \frac{\as^2 G_{\rm F}^2 M^2 \beta_H}{2304(2\pi)^3S}  \Bigg[ \Big|f_{\rm Tri}^{\rm A} + f_{\rm Box}^{\rm A}\Big|^2 + \Big| f_{\rm Box}^{\rm B} \Big|^2 \Bigg] \nno\\
 &\hspace{-2em} \times   \Bigg[ \int_{\sqrt{\tau} e^{-Y}}^{1}\frac{dz}{z}f_{g/A}(\sqrt{\tau}e^{Y},\mu_f)
 f_{g/B}(\frac{\sqrt{\tau}}{z}e^{-Y},\mu_f) \nno \\
 & \hspace{-2em}  +\int_{\sqrt{\tau} e^{Y}}^{1}\frac{dz}{z}f_{g/A}(\frac{\sqrt{\tau}}{z}e^{Y},\mu_f)
 f_{g/B}(\sqrt{\tau}e^{-Y},\mu_f) \Bigg]\frac{C(z,M,\mu_f)}{2}.
\end{align}

\begin{figure}[H]
\begin{center}
\includegraphics[width=0.48\textwidth]{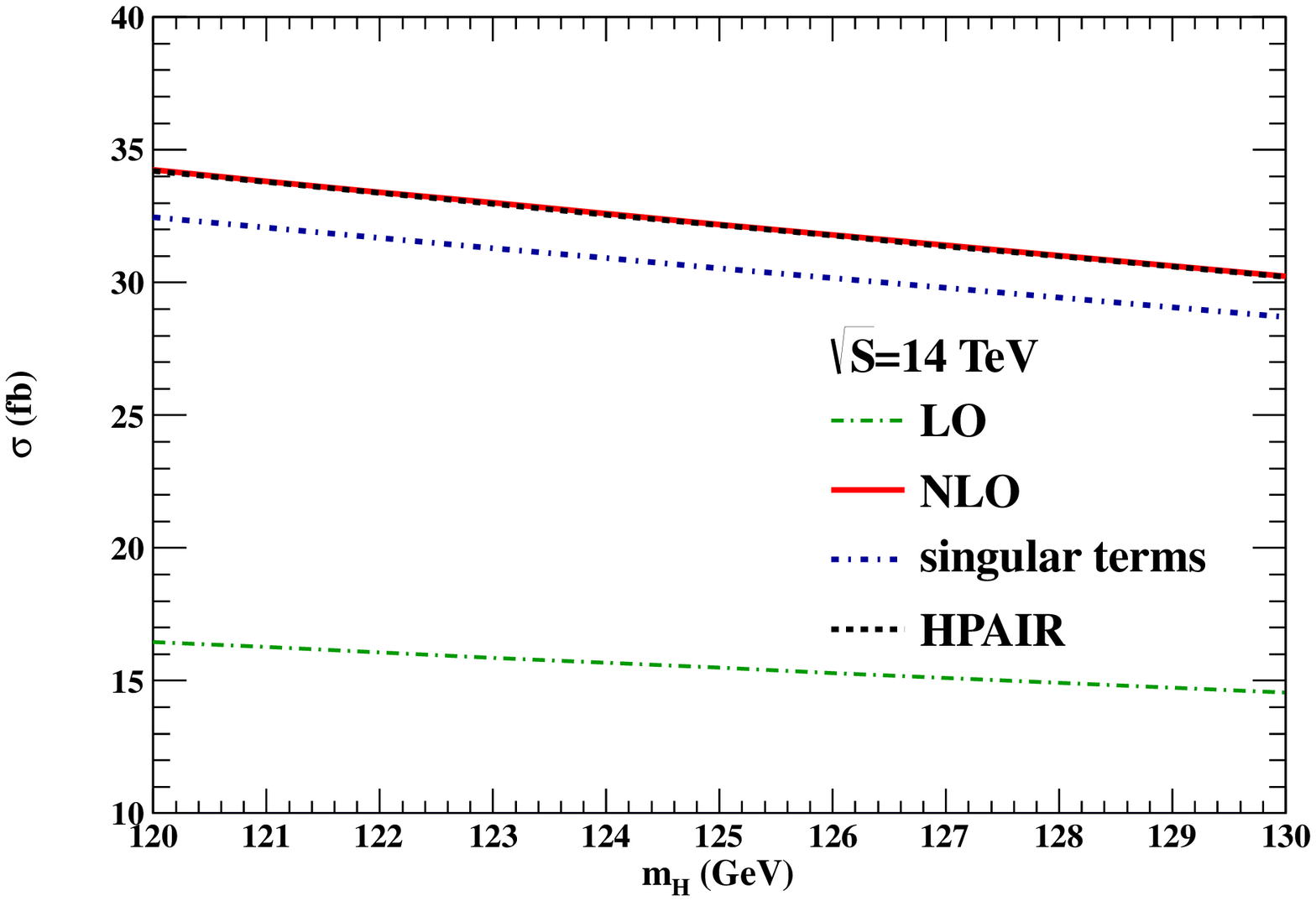}
\quad
\includegraphics[width=0.48\textwidth]{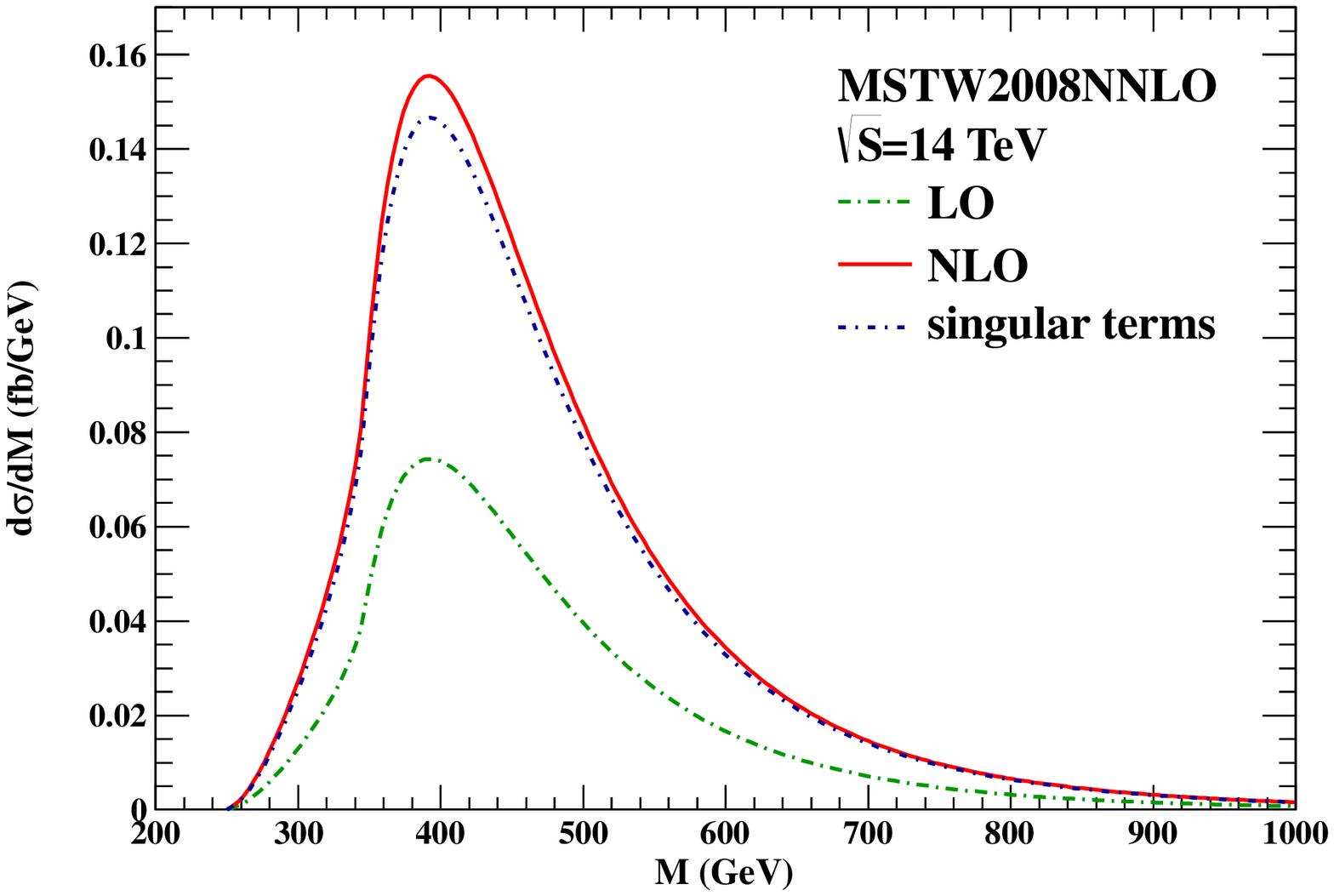}
\end{center}
\vspace{-4mm}
\caption{\label{FO_thres}
Comparison of the NLO results and singular terms contributions to the total cross section (Left) and the invariant mass distributions (Right) for Higgs boson pair production at the LHC with $\sqrt{S}=14$ TeV. The PDF set MSTW2008NNLO and associated strong coupling constant $\as$ are used. The factorization scale is chosen as Higgs boson pair invariant mass. The other SM input parameters are shown in Eq.~(\ref{sminput}).}
\end{figure}

The singular terms~Eq.(\ref{Ckernal}) make the perturbative series badly convergent in the threshold limit $z \to 1$, therefore must be resummed to all orders. However, if the regular terms are comprisable in size to the singular terms, it would make little sense to resum these singular distributions in $(1-z)$. Thus, it is important to compare the contributions from the singular terms and NLO results. In Fig.~\ref{FO_thres} we see that the singular terms contribute $98\%$ of the NLO total cross section and $95\%$ at $M=400$ GeV in Higgs boson pair invariant mass distribution. This suggest that the NLO results are well approximated by the singular terms, which reveals the necessity of resuming these singular terms to all orders. In Fig.~\ref{FO_thres} we also compare our results of the total cross section~(red solid line) and those given by the code \texttt{HPAIR}~(black dashed line) at QCD NLO level~(We have modified the code \texttt{HPAIR} to apply the PDF sets and the associated strong coupling constant $\as$ in LHAPDF library~\cite{Whalley:2005nh}). It is shown that they are well consistent in the range of Monte Carlo integration error.

While this article is in progress, the top quark mass corrections to the total cross section of Higgs boson pair production at the QCD NLO appear in Ref.~\cite{Grigo:2013rya}, in which the NLO cross section containing exact top quark mass effects is expanded in powers of $1/m_t$ and the the power corrections are calculated up to $O(1/m_t^8)$ and $O(1/m_t^{12})$ for partonic channel $gg \to HH$ and $qg(\bar{q}) \to HH$, respectively. They find that the poor convergence induced by top quark mass corrections can be cured if the exact LO cross section are used to normalize the NLO QCD correction and the mass corrections provide $\mathcal{O}(10\%)$ increase in the NLO QCD results. We have compared their approximated methods of normalizing the NLO QCD corrections to the exact LO results and our approximated methods of using form factors in the NLO QCD calculations, and found that these two results agree with each other for the lowest powers of $1/m_t$.

\section{THE HARD AND SOFT FUNCTION AT THE NLO}\label{s3}
Following the approach in Ref.~\cite{Ahrens:2008nc}, the hard scattering kernel $C(z,\mu_f,M)$  in the infinite top quark mass limit can be factorized as
\begin{eqnarray}\label{ck}
 C(z,\mu_f,M) = C_t^2(\mu_f^2) \mathcal{H}(M^2,\mu_f^2) \mathcal{S}\left(s(1-z)^2,\mu_f^2\right),
\end{eqnarray}
where $C_t$ is the Wilson coefficient in Eq.(\ref{Ct}), and it satisfies the RG equation\cite{Ahrens:2008nc}
\begin{eqnarray}
 \frac{d}{d\ln\mu}C_t(\mu^2)=\gamma^{t}(\as)C_t(\mu^2),~~~~~{\rm with}~~ \gamma^{t}(\as) = \as^2\frac{d}{d\as}\frac{\beta(\as)}{\as^2},
\end{eqnarray}
where $\beta(\as)$ is the QCD $\beta$-function. Then we obtain the evolution equation
\begin{eqnarray}
 C_t(\mu_f^2)=\frac{\beta(\as(\mu_f^2))/\as^2(\mu_f^2)}{\beta(\as(\mu_t^2))/\as^2(\mu_t^2)}C_t(\mu_t^2),
\end{eqnarray}
where $\mu_t$ is the matching scale. The hard and soft functions describe interactions at different scales, and they can be calculated order by order in perturbative theory at each scale. At the next-to-next-to-leading logarithmic~(NNLL) accuracy, we need the explicit expressions of hard and soft functions up to NLO.

\subsection{Hard function}
In this process the hard function $\mathcal{H}(M^2,\mu^2)$ is the absolute value squared of the Wilson coefficient
$C_S(-M^2,\mu^2)$~(here and below the negative arguments are understood with a $-i\e$ prescription),
\begin{eqnarray}
 \mathcal{H}(M^2,\mu^2)=\left|C_S(-M^2,\mu^2)\right|^2,
\end{eqnarray}
and $C_S(-M^2,\mu^2)$ can be obtained by matching the two-gluon operators in the full theory onto an operator in SCET, where the infrared divergences are subtracted in the $\overline{\rm MS}$ scheme. We get the Wilson coefficient $C_S(-M^2,\mu^2)$ up to NLO as follows,
\begin{eqnarray}\label{hardcs}
 C_S(-M^2,\mu^2)=1+\frac{\as(\mu^2)}{4\pi}(-3L^2+\frac{\pi^2}{2}),
\end{eqnarray}
where $L=\ln\left(-M^2/\mu^2\right)$, and this result agrees with those in Ref.~\cite{Ahrens:2008nc}. The RG equation for $C_S(-M^2,\mu^2)$ is governed by the anomalous-dimension, the structure of which has been predicted up to four-loop level for the case of massless partons \cite{Ahrens:2012qz}. The $C_S(-M^2,\mu^2)$ satisfies the RG equation
\begin{eqnarray}\label{hardrgeq}
 \frac{d}{d\ln\mu}C_S(-M^2,\mu^2)=
\left[ \Gamma_{\rm cusp}^A(\as)\ln\frac{-M^2}{\mu^2} + \gamma^S(\as)\right]C_S(-M^2,\mu^2),
\end{eqnarray}
where $\Gamma_{\rm cusp}^A(\as)$ is the cusp anomalous dimension of Wilson loops with light-like segments~\cite{Korchemskaya:1992je}, while $\gamma^S(\as)$ controls the single-logarithmic evolution. After solving the RG equations, we get the Wilson coefficient $C_S$
\begin{align}
% C_S(-M^2,\mu_f^2)&=&\exp\bigg[ 2S(\mu_h^2,\mu_f^2)
% - a_\Gamma(\mu_h^2, \mu_f^2)\ln\frac{-M^2}{\mu_h^2}\nno\\
% & & - a_{\gamma^S}(\mu_h^2,\mu_f^2) \bigg]C_S(-M^2,\mu_h^2),
 C_S(-M^2,\mu_f^2) = U_{\rm H}(\mu_f^2, \mu_h^2) C_S(-M^2,\mu_h^2).
\end{align}
Here $\mu_h$ is the hard matching scale and the evolution function $U_{\rm H}(\mu_f^2, \mu_h^2)$ has the form
\begin{align}
 U_{\rm H}(\mu_f^2, \mu_h^2) = \exp\bigg[ 2S(\mu_h^2,\mu_f^2)
 - a_\Gamma(\mu_h^2, \mu_f^2)\ln\frac{-M^2}{\mu_h^2}
 - a_{\gamma^S}(\mu_h^2,\mu_f^2) \bigg],
\end{align}
where $S(\nu^2,\mu^2)$ and $a_\Gamma(\nu^2,\mu^2)$ are defined as
\begin{eqnarray}
 S(\nu^2,\mu^2)&=&-\int_{\as(\nu^2)}^{\as(\mu^2)}d\alpha \frac{\Gamma_{\rm cusp}^A(\alpha)}{\beta(\alpha)}
 \int_{\as(\nu^2)}^{\alpha}\frac{d\alpha'}{\beta(\alpha')}, \\
 a_\Gamma(\nu^2,\mu^2)&=&-\int_{\as(\nu^2)}^{\as(\mu^2)}d\alpha \frac{\Gamma_{\rm cusp}^A(\alpha)}{\beta(\alpha)}.
\end{eqnarray}
$a_{\gamma^S}$ has a similar expression. Up to NNLL level, we need $3$-loop cusp anomalous dimension and $2$-loop normal anomalous dimension, and their explicit expressions are collected in the Appendices of Refs.\cite{Becher:2007ty,Ahrens:2008nc}.

In Eq.(\ref{hardcs}), if we set the hard matching scale $\mu_h^2 \sim M^2$, the perturbative behavior of FO hard matching coefficient is bad due to the existence of $\pi^2$ terms from the squared logarithmic term. This poor perturbative behavior can be avoided if we set the hard matching scale at the time-like region $\mu_h^2 \sim -M^2$. In order to get the hard matching coefficient at the time-like region, we apply the solution of RG Eq.~(\ref{hardrgeq}) to evolve the hard matching coefficient from the space-like region to time-like region. As a result, the $\pi^2-$enhanced terms are resummed to all orders~\cite{Ahrens:2008qu}. In the process of calculation we need to apply the strong coupling $\as(\mu^2)$ evaluated at the negative arguments. Up to NLO, it can be related to the running couplings at the positive argument as
\begin{align}{\label{mmuas}}
 \frac{\as(\nu^2)}{\as(-\mu^2)} = 1 - i \, a(\nu^2) + \frac{\as(\nu^2)}{4\pi}\left[ \frac{\beta_1}{\beta_0}\ln\left[1 - i \, a(\nu^2)\right] + \beta_0 \ln\frac{\mu^2}{\nu^2} \frac{1}{1 - i \, a(\nu^2)} \right] + \mathcal{O}(\as^2),
\end{align}
where $a(\nu^2) = \beta_0\as(\nu^2)/4$, and $\nu^2>0$ is a optional scale far away from Landau pole. In our case, we choose $\nu = \mu_f$. Hence, up to $\mathcal{O}(\as)$ the evolution function $U_{\rm H}$ can be written as
\begin{align}{\label{pisqU_1}}
\ln U_{\rm H}
%  = & \, 2 \mathcal{R}e\left[ 2S(\mu_h^2,\mu_f^2) - a_{\gamma^{S}}(\mu_h^2,\mu_f^2) - a_{\Gamma}(\mu_h^2,\mu_f^2)\ln\frac{M^2}{|\mu_h^2|}\right] \nno \\
  = & \, \frac{\Gamma_0^{\rm A}}{2 \beta_0^2} \Bigg\{ \frac{4\pi}{\as(\mu_f^2)} \left[ 2 a \arctan(a) - \ln(1+a^2) \right]  - \frac{\ln(1+a^2)}{1+a^2} \beta_0 \ln \frac{|\mu_h^2|}{\mu_f^2} \nno \\
 & + \left[ \frac{\Gamma_1^{\rm A}}{\Gamma_0^{\rm A}} - \frac{\beta_1}{\beta_0} - \frac{\gamma_0^{\rm S} \, \beta_0}{\Gamma_0^{\rm A}} - \beta_0 \ln \frac{M^2}{|\mu_h^2|} \right] \ln(1+a^2)  \nno \\
 & + \frac{\beta_1}{4\beta_0} \left[ 4 \arctan^2(a) - \ln^2(1+a^2) \right] -2\beta_0 \frac{a \arctan(a)}{1+a^2} \ln\frac{|\mu_h^2|}{\mu_f^2} \Bigg\} \nno \\
 & + \mathcal{O}(\as).
\end{align}
where $a=a(\mu_f^2)$. This expression have the same form in the case of single Higgs boson production~\cite{Ahrens:2008nc,Ahrens:2008qu} if we set $\mu_f^2=M^2=|\mu_h^2|$. In our predictions the factorization scale is chosen as the invariant mass of Higgs boson pair $\mu_f = M$. Due to the fact that the variable $a(\mu_f^2)<a(4 m_{\rm H}^2)\approx0.195$ can be count as $\mathcal{O}(\as)$, Eq.~(\ref{pisqU_1}) can be simplified as
\begin{align}{\label{pisqU_2}}
 \ln U_{\rm H} = & \frac{\pi C_A}{2}\as(\mu_f^2)\left[ 1 + \frac{\as(\mu_f^2)}{4\pi}\left( \frac{\Gamma_1^{\rm A}}{\Gamma_0^{\rm A}} - 3\beta_0\ln\frac{|\mu_h|^2}{\mu_f^2} - \beta_0 \ln\frac{M^2}{|\mu_h|^2} \right) + \mathcal{O}(\as) \right].
\end{align}
Thus, the $\pi^2$-enhanced correction has the form of $\exp[\pi C_A \as(M^2)/2]$ at the LO, and it decreases with the increasing the invariant mass of Higgs boson pair, which will be numerically discussed in Sec.~\ref{s4}.

\subsection{Soft function}
The soft function $\mathcal{S}(s(1-z)^2,\mu^2)$, describing soft interactions between all colored particles, can be defined as
\begin{eqnarray}
 \mathcal{S}(s(1-z)^2,\mu^2)=\sqrt{s}W(s(1-z)^2,\mu^2).
\end{eqnarray}
In order to derive the RG equation satisfied by the soft function, we employ the fact that the physical cross section in the threshold region is independent of the arbitrary scale $\mu$. We arrive at the integro-differential evolution equation for the momentum-space Wilson coefficient
\begin{eqnarray}
 \frac{d}{d\ln\mu}W(\omega^2,\mu^2)&=&-[4\Gamma_{\rm cusp}(\as)\ln\frac{\omega}{\mu}
 + 2 \gamma^W(\as)]W(\omega^2,\mu^2) \nno\\
 && -4\Gamma_{\rm cusp}(\as)\int_0^{\omega}d \omega'\frac{W(\omega'^2,\mu^2)-W(\omega^2,\mu^2)}{\omega-\omega'},
\end{eqnarray}
where the anomalous dimensions is
\begin{eqnarray}
 \gamma^W=\frac{\beta(\as)}{\as}+\gamma^t+\gamma^S+2\gamma^B,
\end{eqnarray}
in which $\gamma^B$ is the coefficient of $\delta(1-z)$ in the Altaralli-Parisi splitting function $P_{gg}(z)$\cite{Becher:2007ty}. The solution of this evolution equation is\cite{Ahrens:2008nc}
\begin{eqnarray}\label{softevolution}
 \omega W(\omega^2,\mu_f^2)=\exp\left[-4S(\mu_s^2,\mu_f^2) + 2 a_{\gamma^W}(\mu_s^2,\mu_f^2)\right]
 \tilde{s}(\partial_\eta,\mu_s^2)\left(\frac{\omega^2}{\mu_s^2}\right)^{\eta}\frac{e^{-2\gamma_E\eta}}{\Gamma(2\eta)},
\end{eqnarray}
where $\eta=2a_\Gamma(\mu_s^2,\mu_f^2)$, and $\tilde{s}(L_s,\mu^2_s)$ is the Laplace transformation of the soft Wilson loop at the matching scale $\mu_s$. At the NLO it is given by
\begin{eqnarray}
 \tilde{s}(L_s,\mu_s^2)=1+\frac{\as(\mu_s^2)}{4\pi}(6L_s^2+\pi^2).
\end{eqnarray}
Eq.~(\ref{softevolution}) is well defined when $\eta>0$. When $\eta<0$ the results should be understood with the pole contribution subtracted, and we deal with it following the approach in Ref.~\cite{Becher:2007ty}.

\subsection{Final RG improved differential cross section}

After combining the hard and soft function together, we obtain the RG-improved expression for
Eq.(\ref{Ckernal}),
\begin{eqnarray}\label{Cresum}
 C(z,m_t,M,\mu_f)&=&\left[C_t(m_t^2,\mu_t^2)\right]^2\left|C_S(-M^2,\mu_h^2)\right|^2
 U(M^2,\mu_t^2,\mu_h^2,\mu_s^2,\mu_f^2)\frac{z^{-\eta}}{(1-z)^{1-2\eta}} \nno\\
 &&\times \tilde{s}\left(\ln\frac{M^2(1-z)^2}{\mu_s^2z}+\partial_\eta,\mu_s^2\right)\frac{e^{-2\gamma_E\eta}}{\Gamma(2\eta)},
\end{eqnarray}
where
\begin{eqnarray}
 U(M^2,\mu_t^2,\mu_h^2,\mu_s^2,\mu_f^2)&=&\frac{\as^2(\mu_s^2)}{\as^2(\mu_f^2)}
 \left[\frac{\beta(\as(\mu_s^2))/\as^2(\mu_s^2)}{\beta(\as(\mu_t^2))/\as^2(\mu_t^2)}\right]^2
 \left|\left(\frac{-M^2}{\mu_h^2}\right)^{-2a_\Gamma(\mu_h^2,\mu_s^2)}\right| \nno\\
 &&\left|\exp\left[4S(\mu_h^2,\mu_s^2)-2a_{\gamma^S}(\mu_h^2,\mu_s^2)+4a_{\gamma^B}(\mu_s^2,\mu_f^2)\right]\right|.
\end{eqnarray}

In the above expression, if we set scales $\mu_t=\mu_h=\mu_s=\mu_f$, then we can recover the threshold singular terms in Eq.(\ref{Ckernal}), which appear in the FO calculation. Substituting Eq.(\ref{Cresum}) into Eq.(\ref{cs}), we can get the resummed differential cross section at the threshold region.

To give precise predictions, we resum the singular terms to all orders and include the non-singular terms up to NLO. The RG-improved differential cross section can be written as
\begin{eqnarray}
 \frac{d^2\sigma^{\rm NNLL+NLO}}{dM^2dY} = \frac{d^2\sigma^{\rm NNLL}}{dM^2dY} +
 \left(\frac{d^2\sigma^{\rm NLO}}{dM^2dY} - \frac{d^2\sigma^{\rm NNLL}}{dM^2dY}\right)\Bigg|_{\rm expanded ~ to ~ NLO}.
\end{eqnarray}
In the threshold region, the second term in the above expression vanishes, and only the threshold contribution dominates, and in the region far away from the threshold region, the resummation effect is not important and the differential cross section is dominated by the NLO results.

\section{NUMERICAL DISCUSSIONS}\label{s4}

In this section, we discuss the numerical results for the threshold resummation effects on Higgs boson pair production at the LHC. We choose the following SM input parameters
\begin{eqnarray}\label{sminput}
 &&G_F=1.166379\times10^{-5}~{\rm GeV}^{-2}, ~~~~m_t=173.2~{\rm GeV} \nno \\
 &&m_H=125~{\rm GeV}, ~~~~\lambda=\lambda_{\rm SM}.
\end{eqnarray}
\noindent Throughout the numerical calculations, we use the MSTW2008NNLO PDF sets and associated strong coupling constant $\as$ unless specified otherwise. The factorization scale $\mu_f$ is choose as the Higgs boson pair invariant mass $M$ unless specified otherwise. There are still three matching scales, i.e. $\mu_t$, $\mu_h$, $\mu_s$, which should be chosen such that the Wilson coefficients evaluated at the matching scales have stable perturbative expansions.

\subsection{Scale setting and scale uncertainties}
As is shown in Eq.~(\ref{hierachy}), the scale hierarchy is assumed in the derivation of the Wilson coefficient $C_t(\mu^2)$, and the effective Lagrange Eq.~(\ref{effL}) is constructed when top quark loop is integrated out. Hence, it is natural to choose the matching scale as $\mu_t=m_t$\cite{Ahrens:2008nc}.

\begin{figure}
 \begin{center}
  \includegraphics[width=0.48\textwidth]{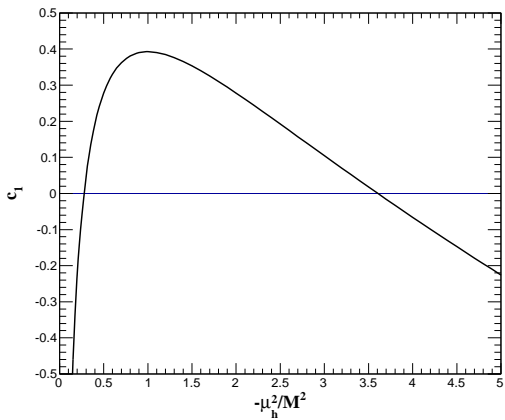}\\
  \caption{Dependence of $c_1$ on the hard matching scale squared $\mu_h^2$.}\label{choosemuh}
  \end{center}
\end{figure}

The hard matching coefficient can be expanded in the perturbative series as
\begin{eqnarray}
 C_S(-M^2,\mu_h^2)=1+\sum_{n=1}^{\infty}c_n\left(\frac{\mu_h^2}{-M^2}\right)\as^n(\mu_h^2).
\end{eqnarray}
In Fig.~\ref{choosemuh} the dependence of $c_1$ on the hard matching scale squared $\mu_h^2$ is shown. The coefficient $c_1$ vanishes for $-\mu_h^2 \sim 3.6M^2$ and $-\mu_h^2 \sim 0.28M^2$. Around the region of second solution the coefficient $c_1$ varies strongly, thus we discard it. In our numerical results, we will take $-\mu_h^2=3.6M^2$ as the default choice. Thus the $\pi^2$-enhanced corrections are contained into the evolution function $U$.

\begin{figure}
\begin{center}
\includegraphics[width=0.48\textwidth]{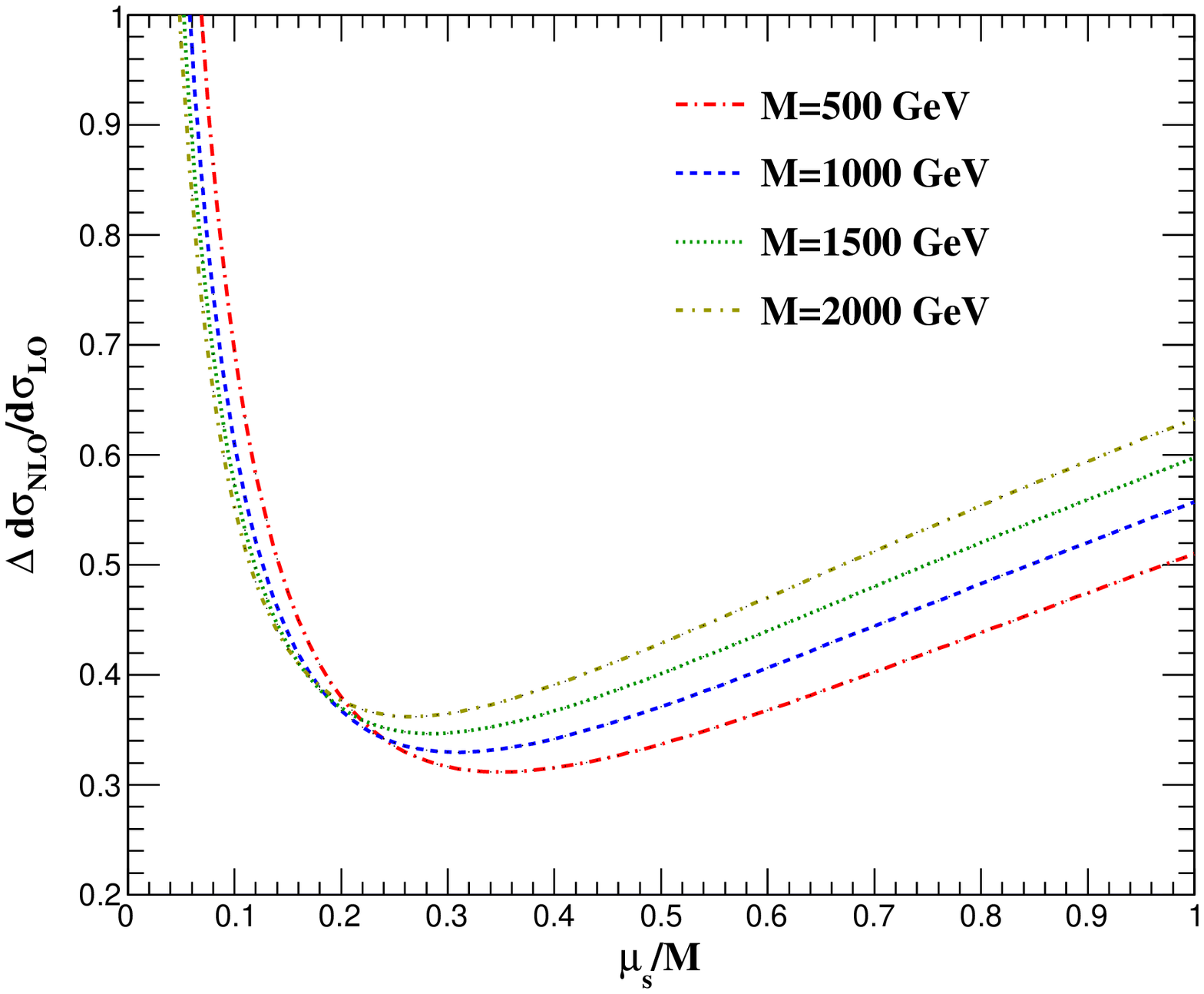}
\quad
\includegraphics[width=0.48\textwidth]{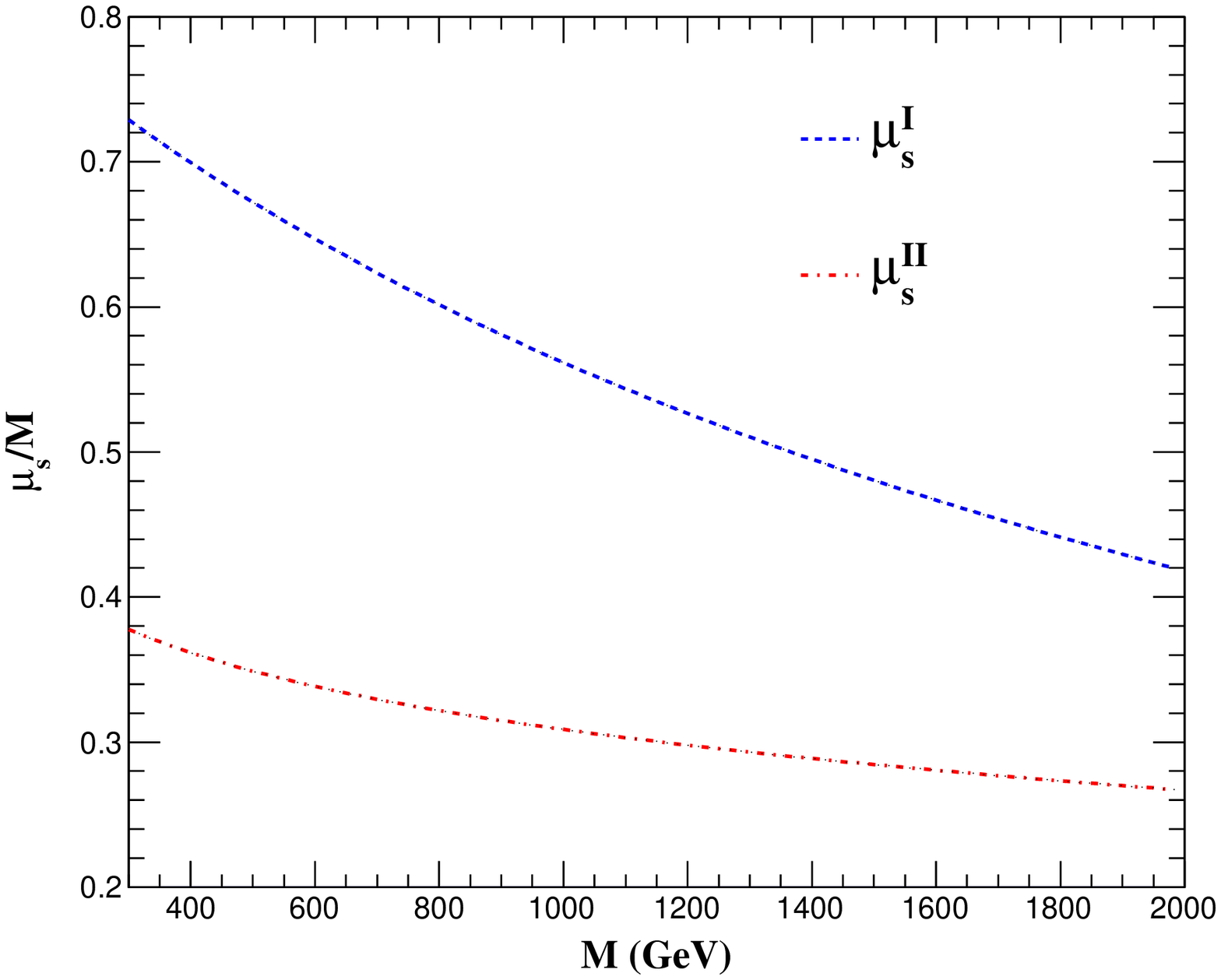}
\end{center}
\vspace{-4mm}
\caption{\label{choosemus}
Left: The contributions of the one-loop corrections to the soft functions to the differential cross section $d\sigma/d M$ for the different values of $M$ at the LHC with $\sqrt{S} = 14$~TeV. Right: Results for the soft matching scale $\mu_s$ for different $M$.}
\end{figure}

For the soft matching scale $\mu_s$, the situations are not so clear, since the soft function $\mathcal{S}(s(1-z)^2,\mu^2)$ depends on the variable $z$. The soft scale is chosen such that the perturbative expression of the soft function have a well-behaved convergence after performing the integration over the variable $z$. In Fig~\ref{choosemus}, we present the corrections from soft functions at the NLO at the LHC with $\sqrt{S} = 14$~TeV. The plots for the LHC with $\sqrt{S} = 33$~TeV are similarly and not shown here. Following the methods in Refs.~\cite{Ahrens:2008nc,Becher:2007ty}, we determine the soft matching scale in two different schemes as is shown below,
\begin{description}
 \item[I].~ Starting from a high scale, choose the value of $\mu_s$ where the soft correction drops below 40\%;
 \item[II].~ Choose the value of $\mu_s$ where the soft correction is minimized.
\end{description}
Hence the soft scales are shown in the right plot of Fig.~\ref{choosemus}, and they can be parameterized by the functions $\mu_s^{\rm I}$ and $\mu_s^{\rm II}$
\begin{eqnarray}
\mu_s^{\rm I}= \frac{M(1-\tau)}{a^{\rm I} + b^{\rm I} \tau^{1/2}},~~~\mu_s^{\rm II}= \frac{M(1-\tau)}{\sqrt{a^{\rm II} + b^{\rm II} \tau^{1/4}}},
\end{eqnarray}
where $(a^{\rm I}, b^{\rm I}, a^{\rm II}, b^{\rm II}) = (1.2, 8.0, 2.9, 28)$ and $(1.0, 6.5, 3.4, 28)$ for the LHC with $\sqrt{S} = 14$ and $33$~TeV, respectively.

\begin{figure}
\begin{center}
\includegraphics[width=0.48\textwidth]{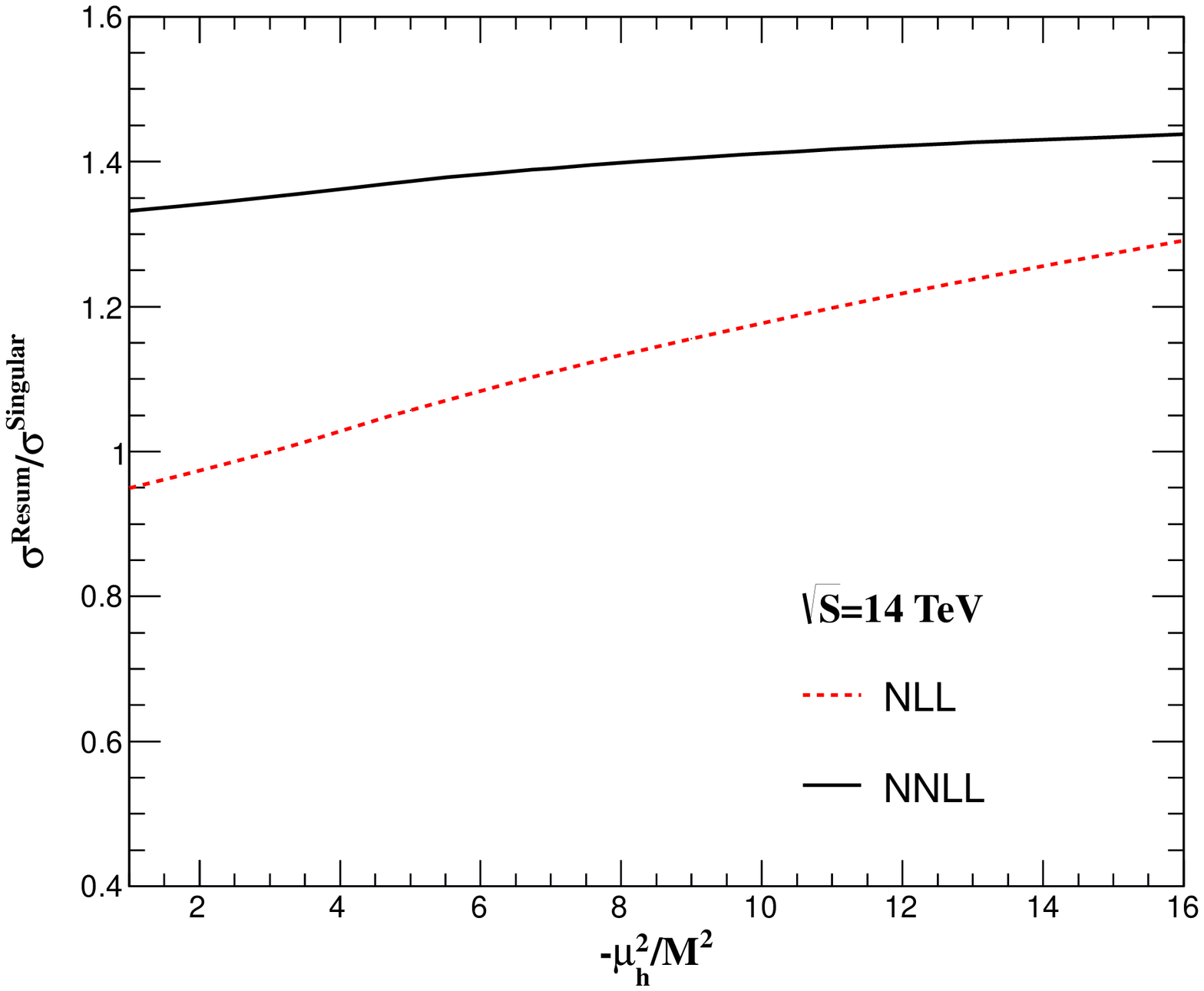}
\quad
\includegraphics[width=0.48\textwidth]{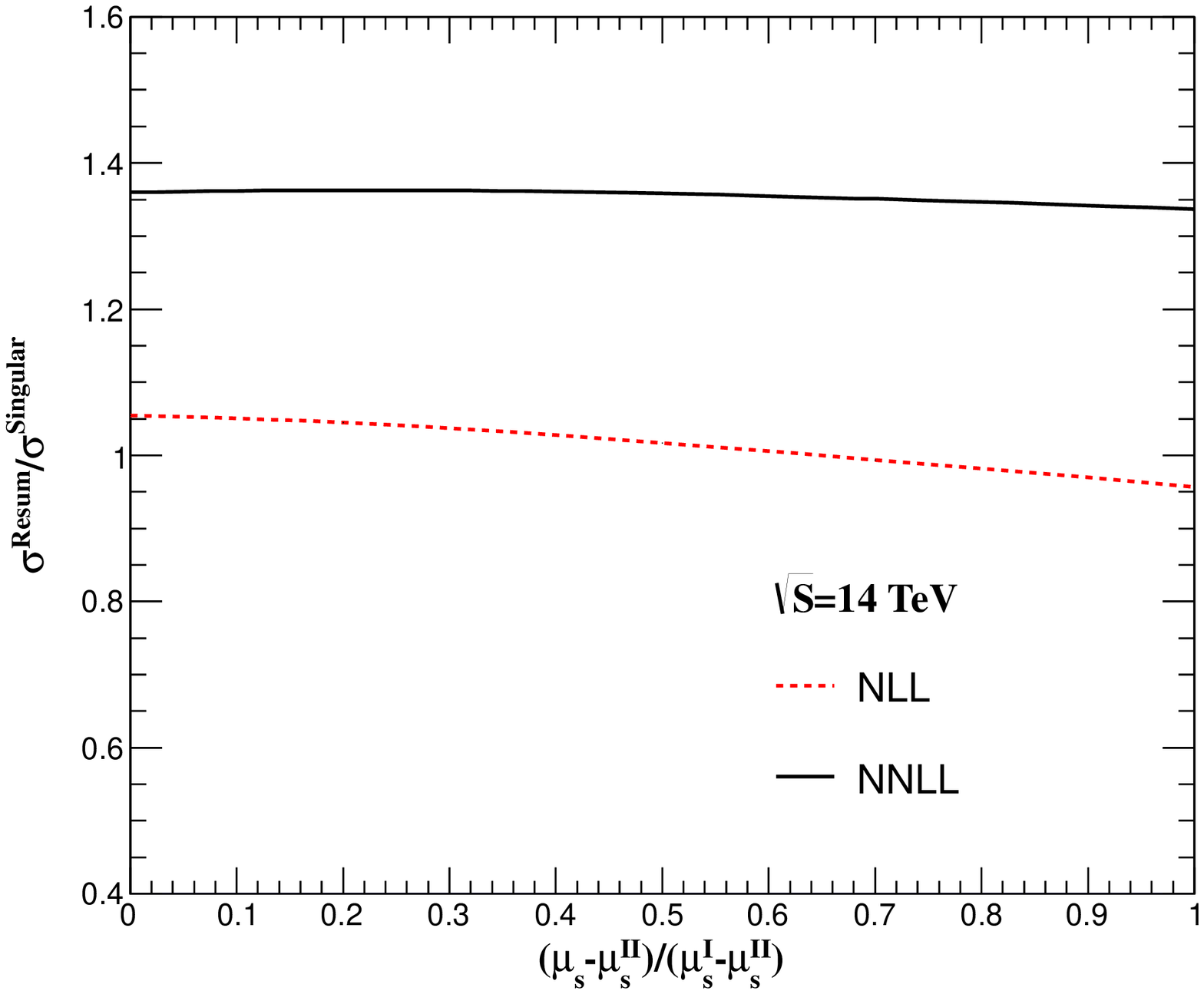}
\end{center}
\vspace{-4mm}
\caption{\label{hs_scale_dep}
The scale dependence of the resummed cross section for the Higgs boson pair production at the next-to-leading logarithmic~(NLL) and NNLL level on the hard matching scale (Left) and the soft matching scale (Right).}
\end{figure}
In Fig.~\ref{hs_scale_dep}, we show the hard and soft scale dependences of the resummed cross section normalized to the singular cross section at the LHC with $\sqrt{S}=14 \, {\rm TeV}$. Here the two matching scales are changed in the region $M^2 < -\mu_h^2 < 16 M^2$ and $\mu_s^{\rm II} < \mu_s < \mu_s^{\rm I}$, respectively. From Fig.~\ref{hs_scale_dep}, it is obvious that the scale dependence is reduced at the NNLL level. The scale dependences on the hard and soft matching scale are about $8\%$ and $2\%$ respectively.

\begin{figure}
\begin{center}
\includegraphics[width=0.48\textwidth]{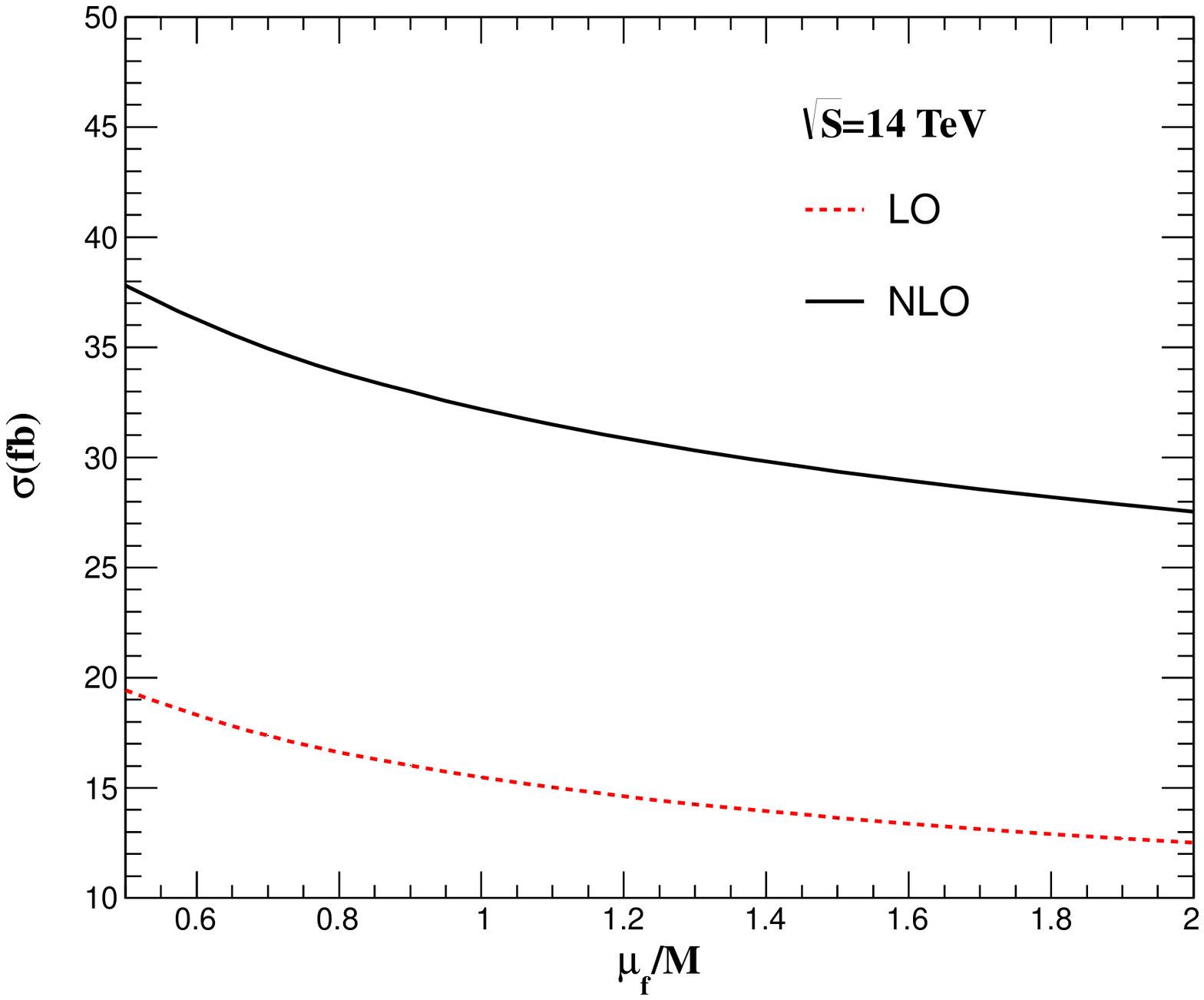}
\quad
\includegraphics[width=0.48\textwidth]{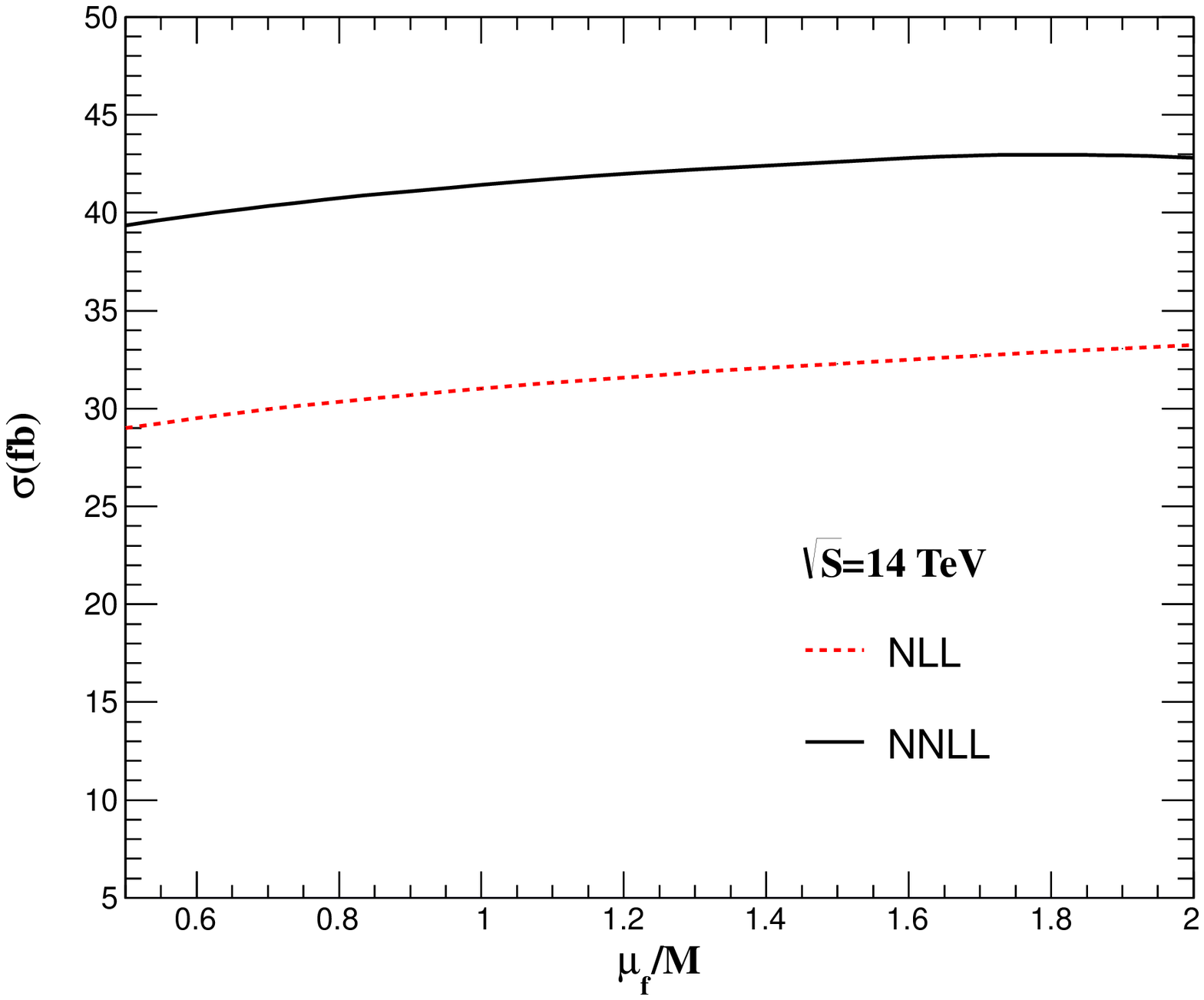}
\end{center}
\vspace{-4mm}
\caption{\label{scale_dep}
The factorization scale dependence of the FO (Left) and resummed (Right) total cross section for the Higgs boson pair production.}
\end{figure}

In Fig.~\ref{scale_dep} we compare the factorization scale dependence of the FO and resumed cross section, where the factorization scale $\mu_f$ is changed from $M/2$ to $2M$. In the FO calculations, the scale dependence is very large, and even at the NLO the factorization scale uncertainty is about 30\%. In contrast the factorization scale dependence is obviously reduced after performing resummation calculations, and the scale uncertainty is only $9\%$ at the NNLL level.

\subsection{Rapidity distributions}
\begin{figure}
\begin{center}
\includegraphics[width=0.48\textwidth]{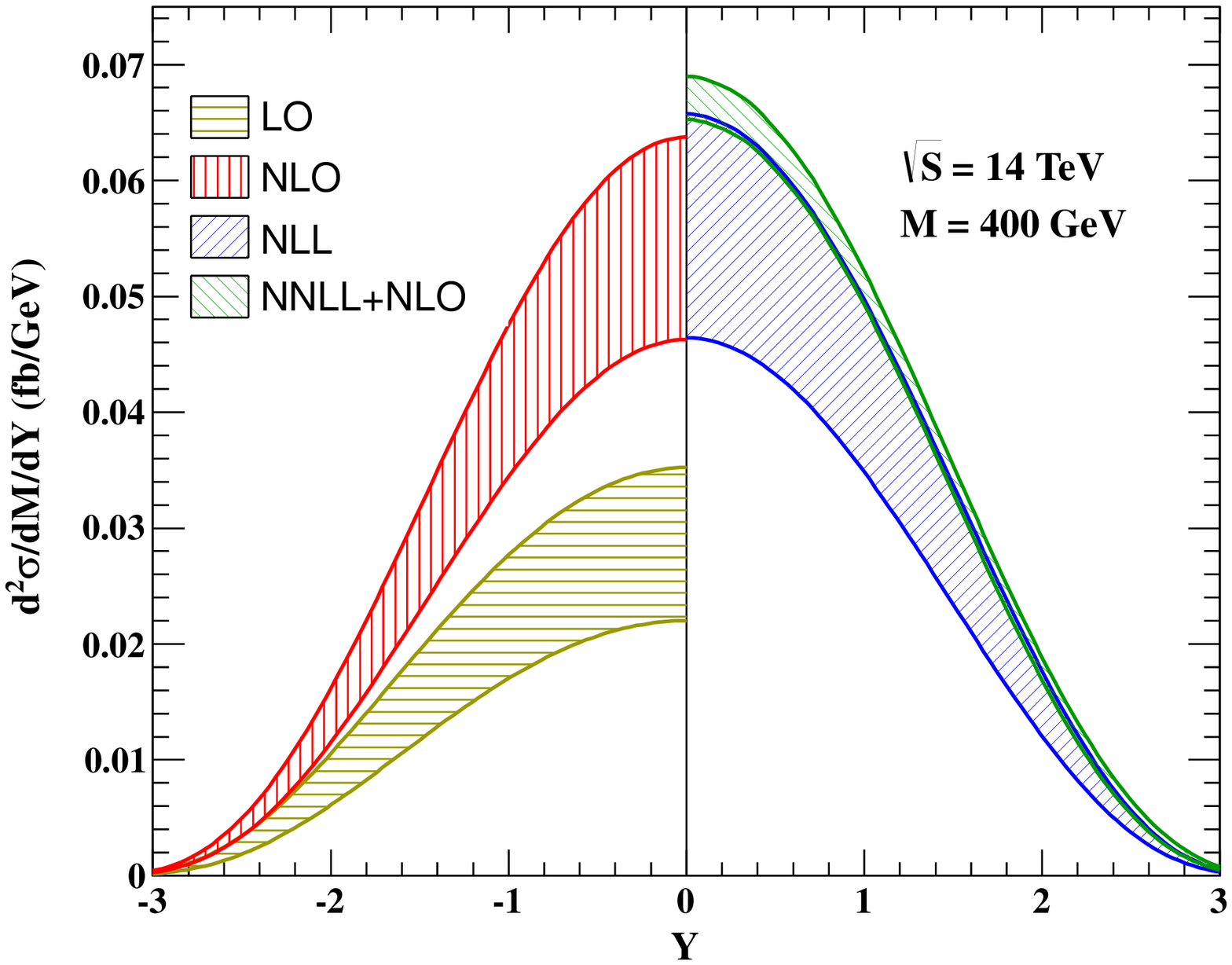}
\quad
\includegraphics[width=0.48\textwidth]{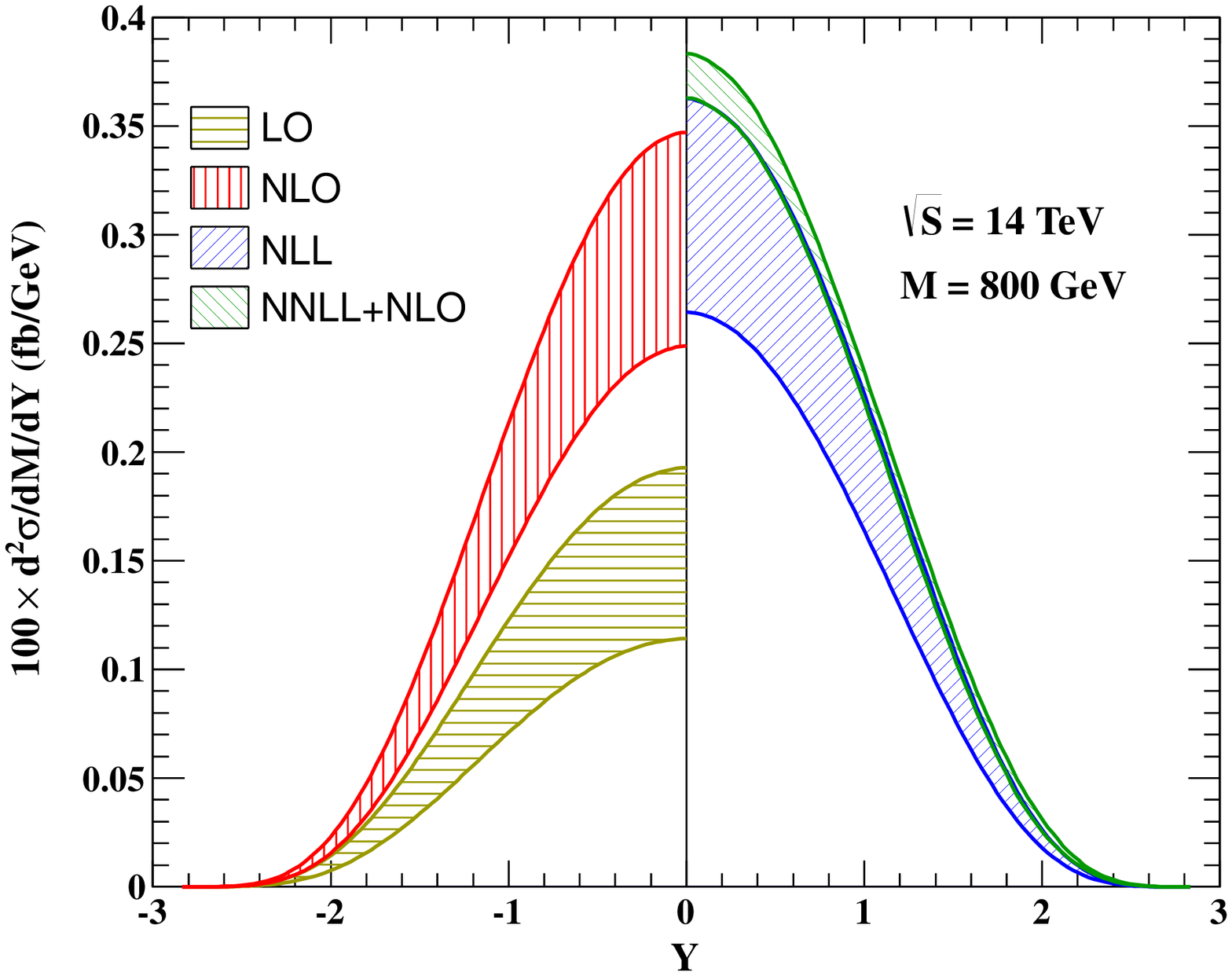}
\end{center}
\vspace{-4mm}
\caption{\label{rap}
Rapidity distributions for Higgs boson pair production at fixed $M=400$ and $800$~GeV with $\sqrt{S}=14~{\rm TeV}$. The bands indicate the scale uncertainties. The yellow and red bands are the FO results at the LO and NLO, respectively. The blue band is the NLL resummation results and the green band includes the effects of NNLL resummation matched to NLO results. }
\end{figure}

In Fig.~\ref{rap}, we show the FO and resummation predictions of rapidity distributions for Higgs boson pair production at $M=400$ and $800$~GeV with $\sqrt{S}=14~{\rm TeV}$, where the bands reflect the scale uncertainties by varying the scales in the range $M/2<\mu_f<2M$, $M^2 < -\mu_h^2 < 16M^2$ and $\mu_s^{\rm II} < \mu_s < \mu_s^{\rm I}$, respectively. In the FO distributions only the factorization scale uncertainties are included, while in the resummation cases these three kinds of uncertainties are added in quadrature. We use MSTW2008LO PDF sets for the LO results, MSTW2008NLO PDF sets for the NLO and NLL results, and MSTW2008NNLO PDF sets for the NNLL+NLO results.

\indent Obviously, the scale uncertainties of FO predictions are not reduced from LO to NLO. However, even there are three kinds of scales uncertainties included in the NNLL+NLO distribution, the scale uncertainties are much smaller than in the FO cases. Moreover, the convergence of the perturbative expansion is also improved after performing resummation. In  Fig.~\ref{rap}, the left plot shows the rapidity distribution of the Higgs boson pair when the invariant mass $M=400$~GeV, and the right plot shows the distribution when $M=800$~GeV. We can see that the rapidity distribution moves toward the central region with the increasing of the invariant mass of the Higgs boson pair.

\subsection{Invariant mass distributions}
\begin{figure}[H]
\begin{center}
\includegraphics[width=0.48\textwidth]{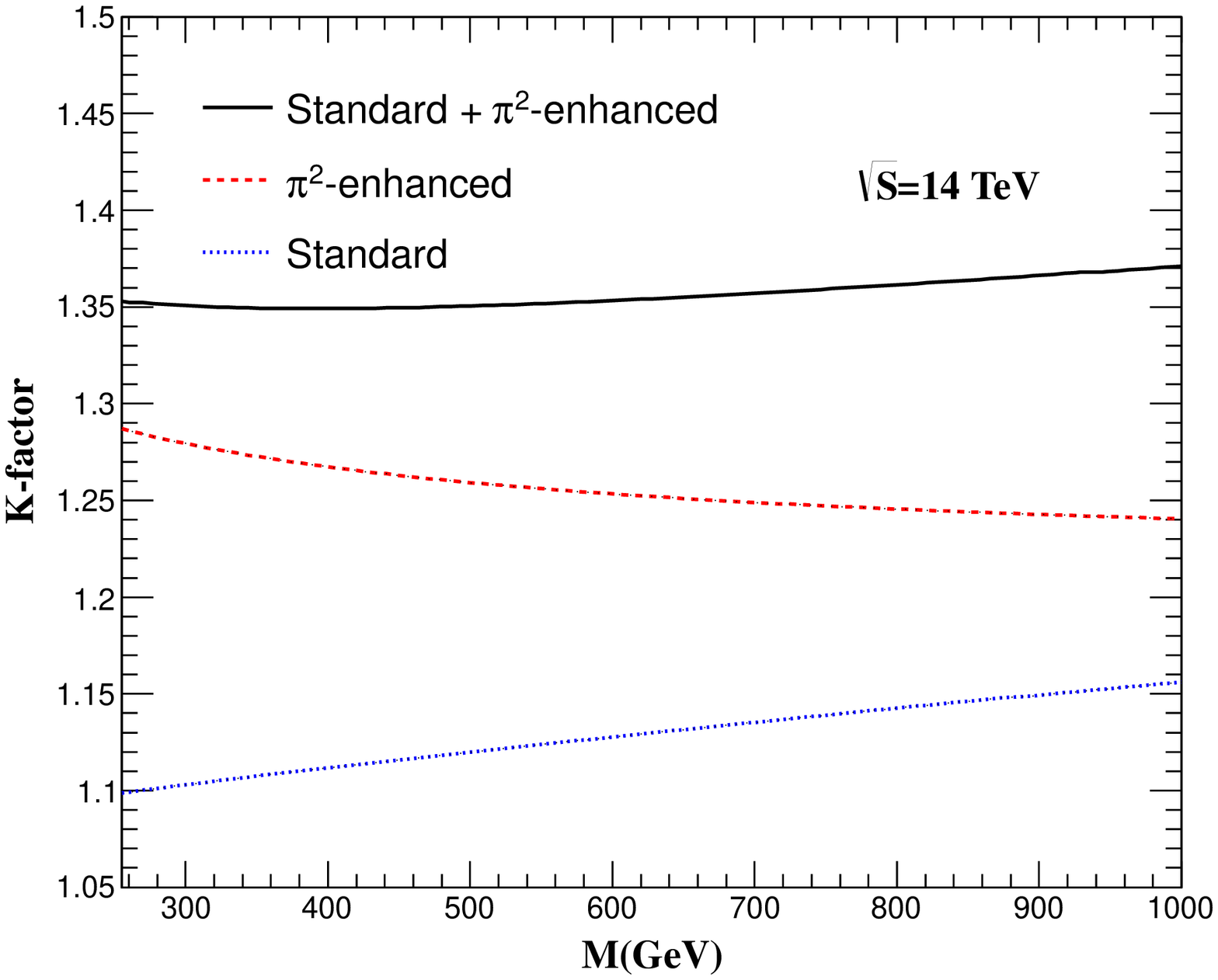}
\quad
\includegraphics[width=0.48\textwidth]{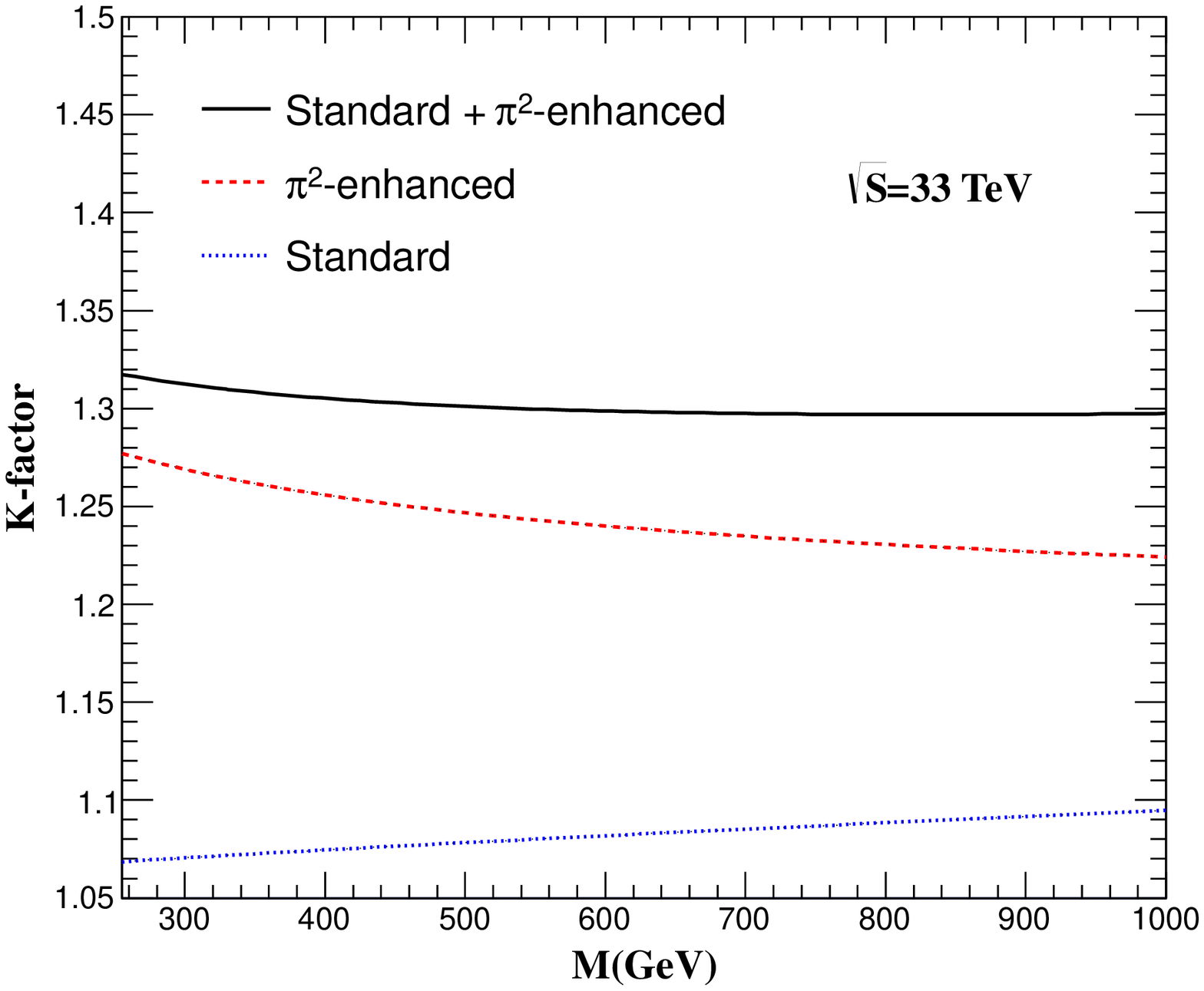}
\end{center}
\vspace{-4mm}
\caption{\label{K_piresum}
 K-factor for Higgs boson pair invariant mass distribution at the LHC with $\sqrt{S}=14$ and $33$~TeV. The black solid line is complete NNLL resummation results. The blue dotted and red dashed line are standard soft-gluon and $\pi^2$ -enhanced terms resummation results, respectively.
 }
\end{figure}

After integrating over the rapidity $\ln\tau/2 \leq Y \leq -\ln\tau/2$, we can get the differential cross section of the invariant mass for the Higgs boson pair. In our calculations the logarithmic terms near the partonic threshold, i.e., standard soft gluon resummation, and the $\pi^2$-enhanced terms from the analytic continuation of the hard function $\mathcal{H}$ to time-like kinematics are both resummed. In order to estimate the relative importance between them, we present the K-factor for Higgs boson pair invariant mass distribution at the LHC with $\sqrt{S}=14$ and $33$~TeV in Fig.~\ref{K_piresum}, where we define the K-factor of the invariant mass distributions as
\begin{eqnarray}
 \frac{d\sigma^{\rm NNLL}}{dM} = K(M) \frac{d\sigma^{\rm Singular}}{dM}.
\end{eqnarray}

When we choose $\mu_h^2=3.6M^2$ the contributions from resummation of $\pi^2$-enhanced terms can be switched off. Only the logarithmic terms Eq.(\ref{Dz}) near the partonic threshold~($z \to 1$) are resummed. In the calculations the hadronic threshold $\tau=M^2/S$ is not very close to 1. However, because the parton luminosity strongly fall off with the increasing the Bjorken variable $x$, the dominance of the partonic threshold region $z \sim 1$ arises dynamically~\cite{Becher:2007ty}. Obviously, in Fig.~\ref{K_piresum} the K-factors corresponding standard soft-gluon resummation effects~(blue dotted line) increase with the increasing of the invariant mass, because gluons radiated from initial states are softer and the logarithmic terms near the partonic threshold are larger in the large invariant mass region than in the small invariant mass region. With the increasing of the center-of-mass energy $\sqrt{S}$ of the collider, this enhanced effects become weak due to the fact that the hadronic threshold $\tau$ is far more less than 1 and the fall-off of the parton luminosity is not much strong.

\begin{figure}
\begin{center}
\includegraphics[width=0.48\textwidth]{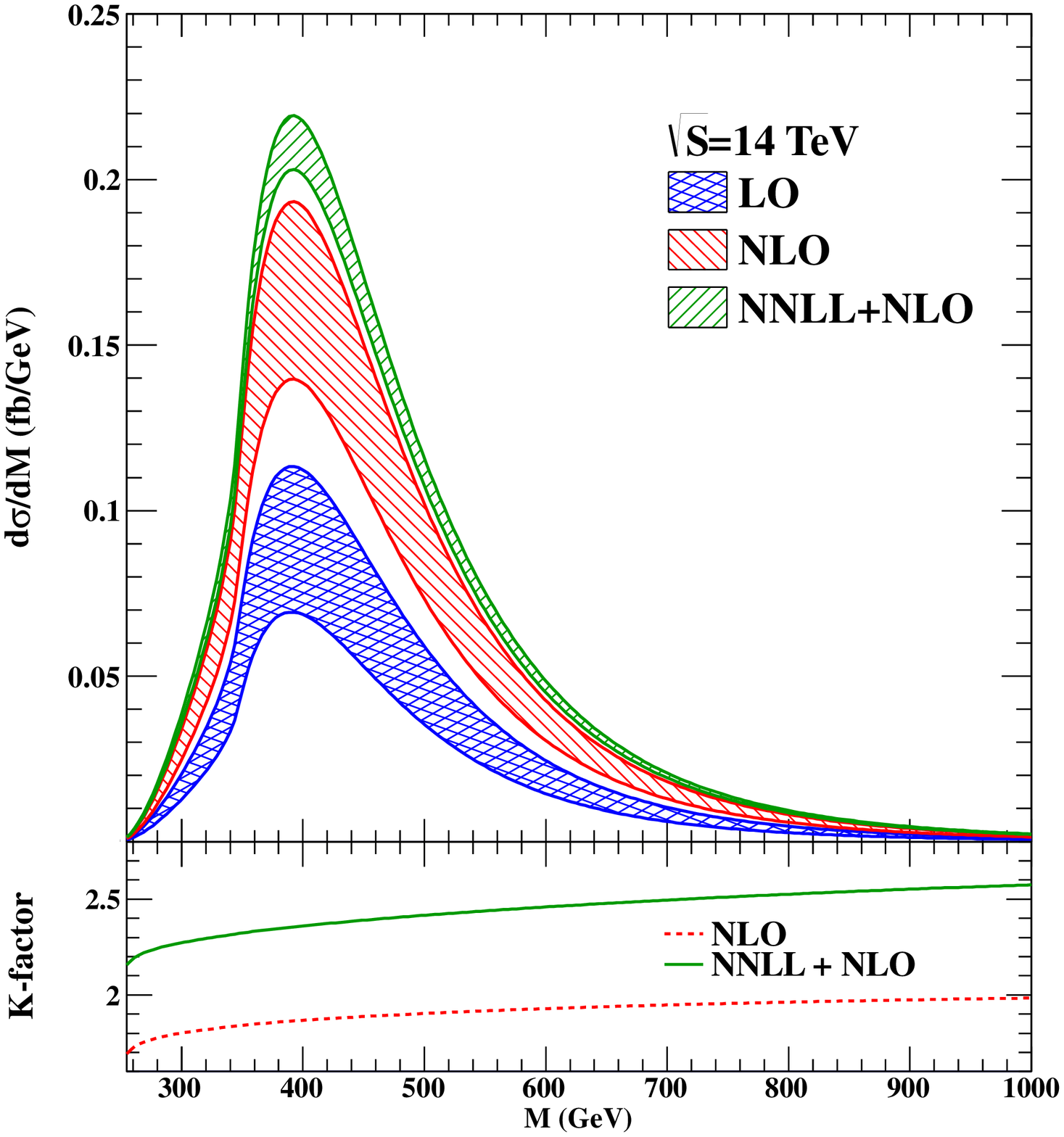}
\quad
\includegraphics[width=0.48\textwidth]{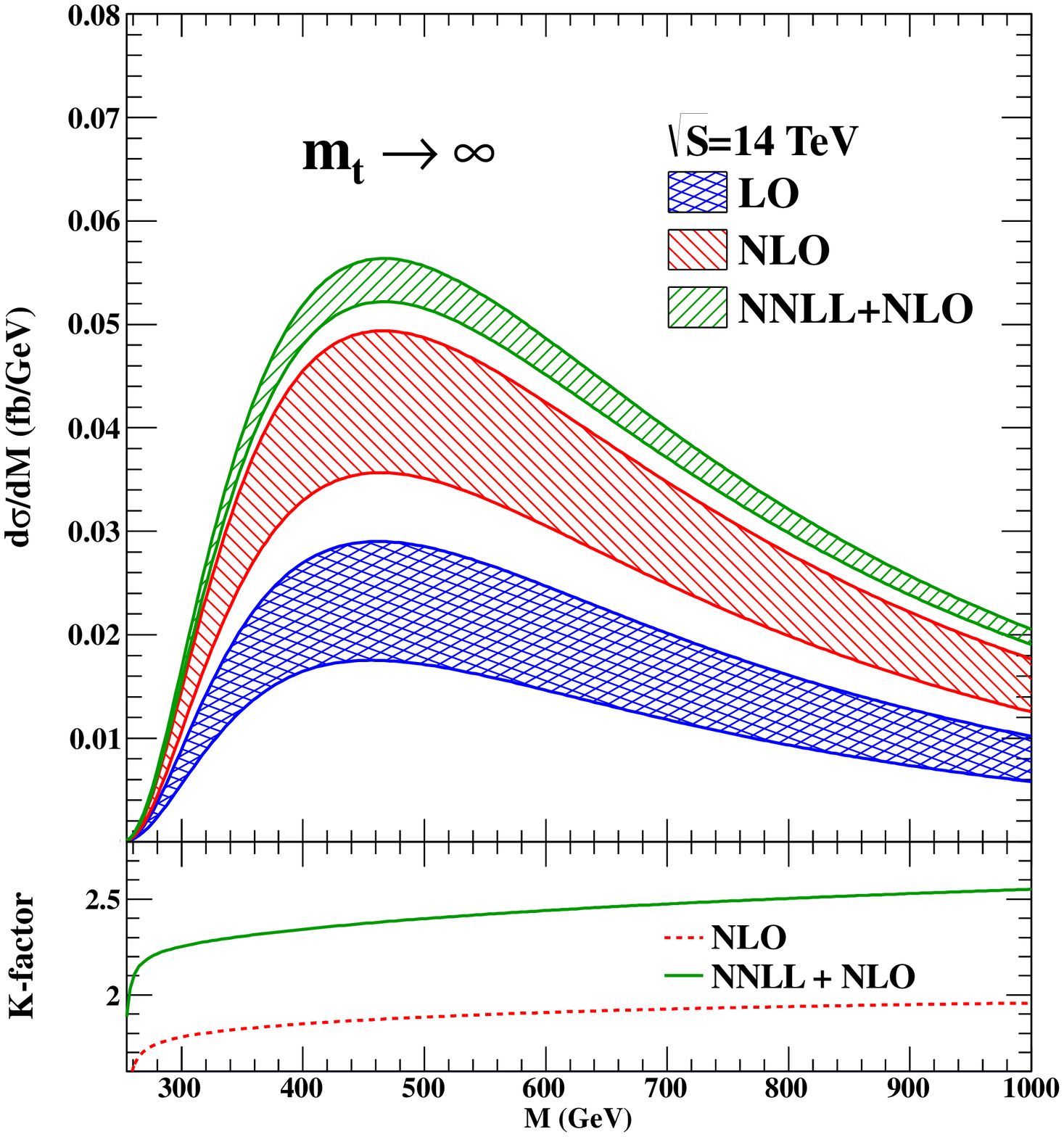}
\quad
\includegraphics[width=0.48\textwidth]{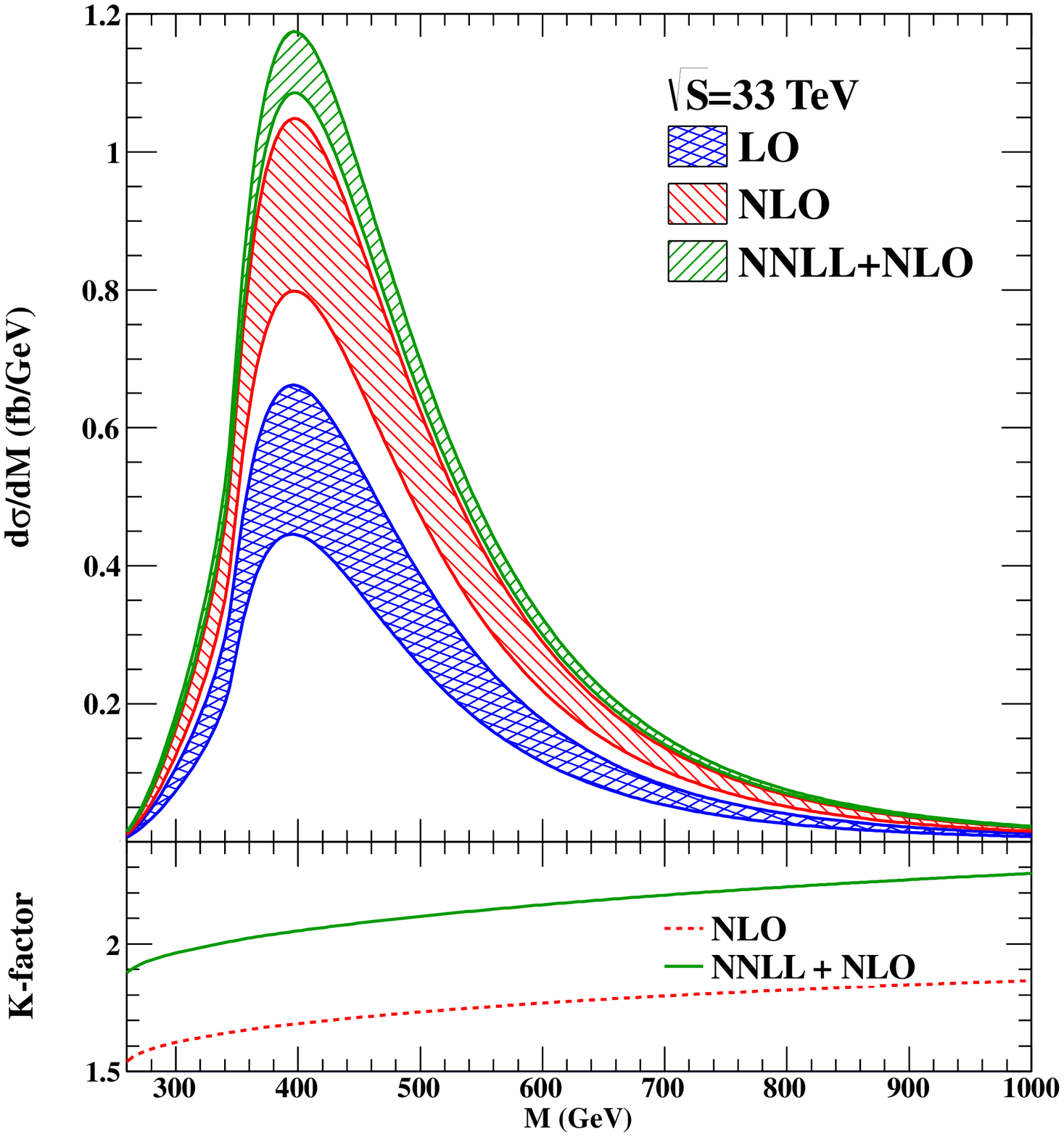}
\quad
\includegraphics[width=0.48\textwidth]{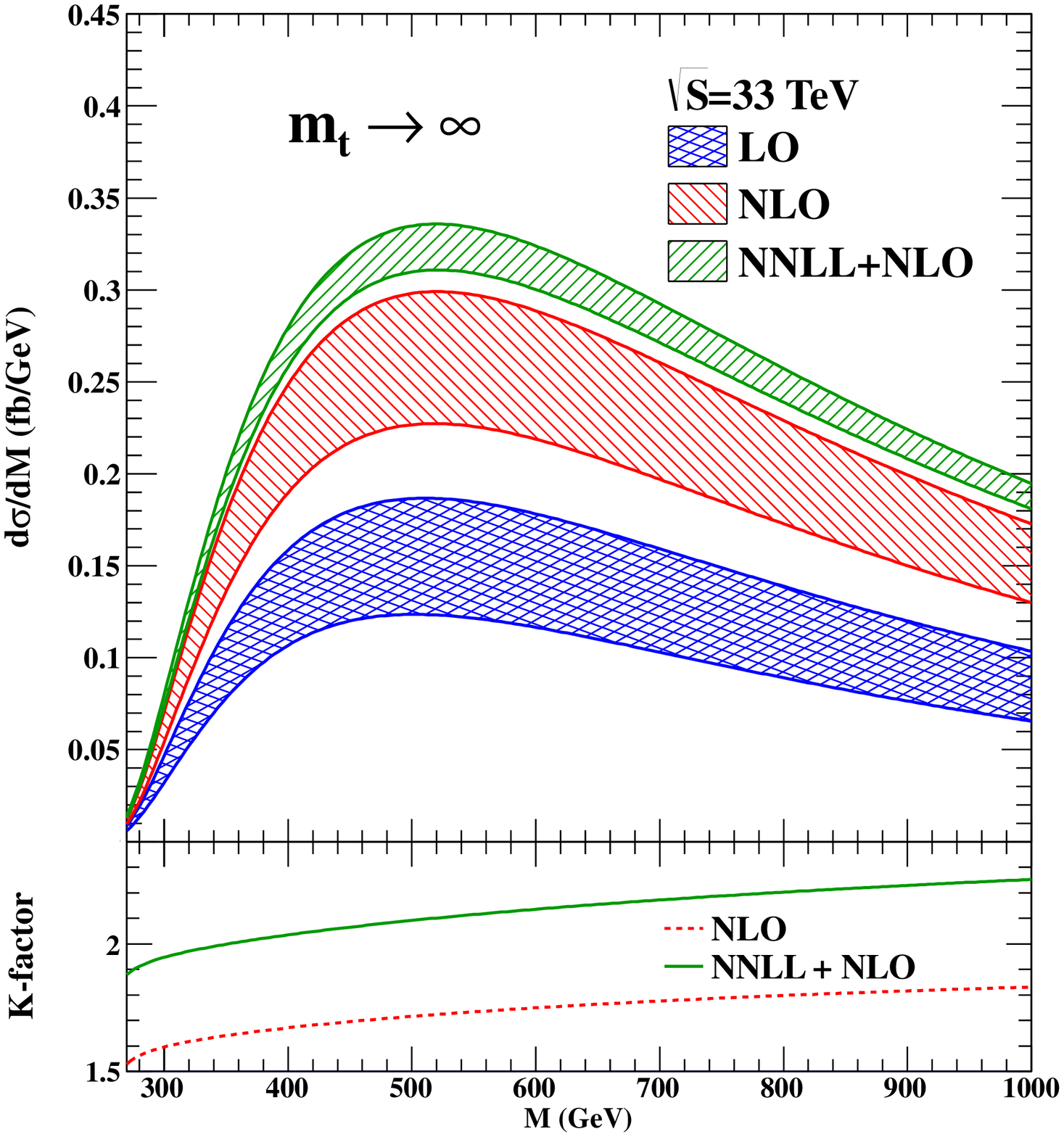}
\end{center}
\vspace{-4mm}
\caption{\label{invmass}
Invariant mass distributions and the associated K-factors for Higgs boson pair production at the LHC with $\sqrt{S}=14$ and $33~{\rm TeV}$. The bands indicate the scale uncertainties. The blue and red bands are FO results at the LO and NLO, respectively, and the green band includes the effects of NNLL resummation matched to NLO results. The green solid and red dashed lines are the K-factors defined as $d\sigma^{\rm NNLL+NLO}/d\sigma^{\rm LO}$ and $d\sigma^{\rm NLO}/d\sigma^{\rm LO}$, respectively.}
\end{figure}

When the hard scale $\mu_h^2=-3.6M^2$ is chosen in the time-like region while the other matching scales are set equal to the factorization scale $\mu_f$, only the contributions from resummation of $\pi^2$-enhanced terms are retained. As is stated in Sec.\ref{s3} the corrections from $\pi^2$-enhanced terms decrease with the increasing the invariant mass of Higgs boson pair, which is obviously shown in Fig.~\ref{K_piresum}~(red dashed line). Besides, these enhanced effects hardly change from $\sqrt{S} = 14~$TeV to $33~$TeV due to the fact that $\pi^2$-enhanced effects do not strongly depend on the the center-of-mass energy of the collider. Combing these two enhanced effects, the K-factors~(black solid lines) almost do not change at both the $14$ and $33$~TeV LHC.

Although our calculations are not sufficient to reproduce all kinematics distribution of the full theory, it is also significant to study the enhancement effects of resummation on the differential invariant mass distribution. In Fig.~\ref{invmass}, we present the Higgs boson pair invariant mass distribution at the LHC with $\sqrt{S}=14~$ and $33~$TeV, where the bands reflect the scale uncertainties associated with the scales $\mu_h$, $\mu_s$ and $\mu_f$. In the left plots we employ the the full expressions of form factors in Eq.~(\ref{formfac1}) including exact top quark mass effects at the LO, while in the right plots the form factors in Eq.~(\ref{formfac_lmt}) in the infinite top quark mass limit are used. In order to compare the enhancement effects of NLO and NNLL+NLO predictions we also present the K-factors~(ratios of NLO and NNLL+NLO to LO, respectively) of the invariant mass distributions in Fig.~\ref{invmass}. In the calculations we use MSTW2008LO, MSTW2008NLO and MSTW2008NNLO PDF sets for the LO, NLO and NNLL+NLO results, respectively. It is obvious that the resummation calculations increase the NLO differential cross section $d\sigma/dM^2$ by 30\%. Moreover, the scale uncertainties are reduced after performing resummation calculations. We also find that the shapes and the heights of the peak for the the invariant mass distributions in the left and right plots are obviously different. Therefore, the infinite top quark limit does not provide reliable predictions for the invariant mass distribution, which have been discussed in Ref.~\cite{Baur:2002rb}.

\begin{figure}
\begin{center}
\includegraphics[width=0.48\textwidth]{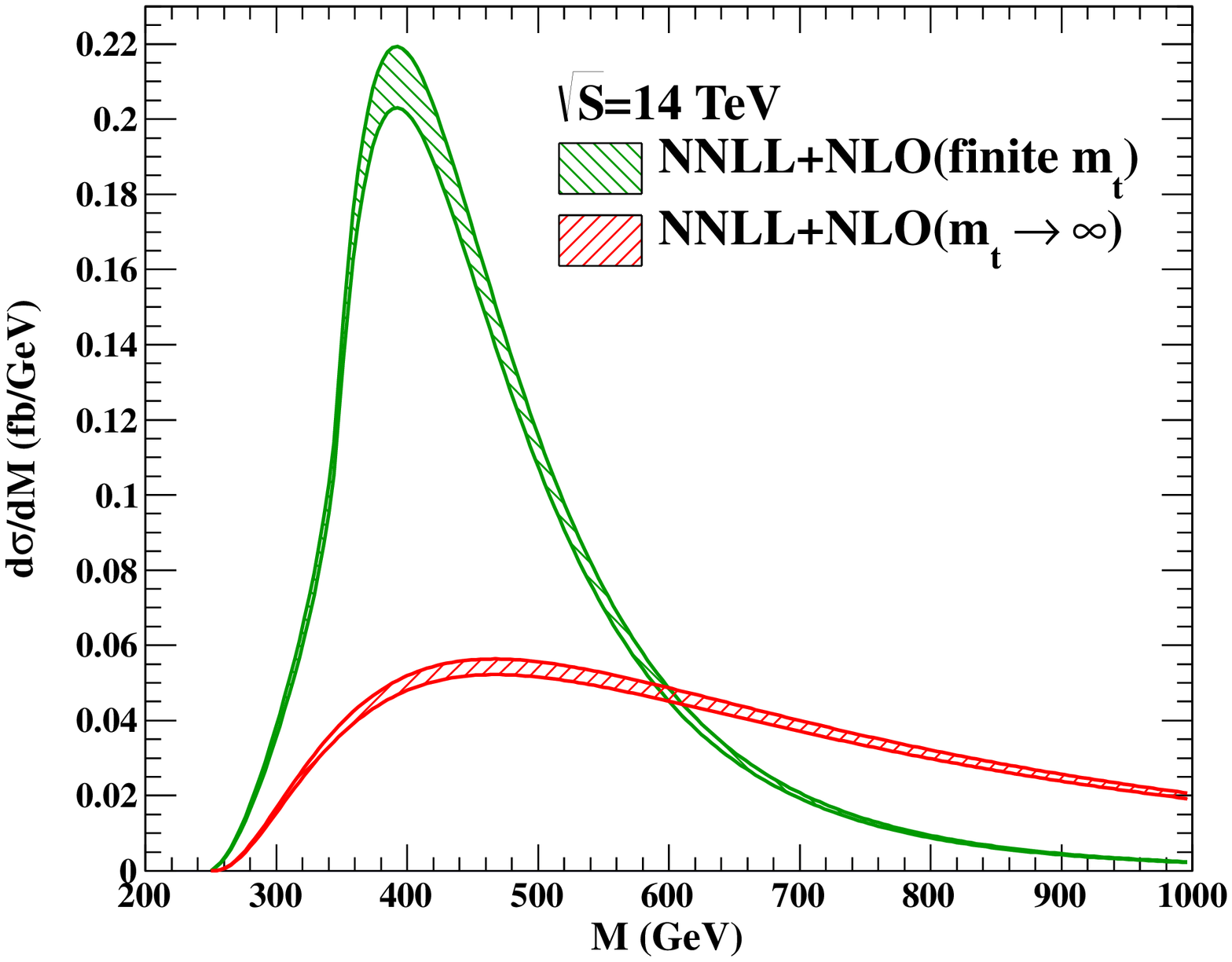}
\quad
\includegraphics[width=0.48\textwidth]{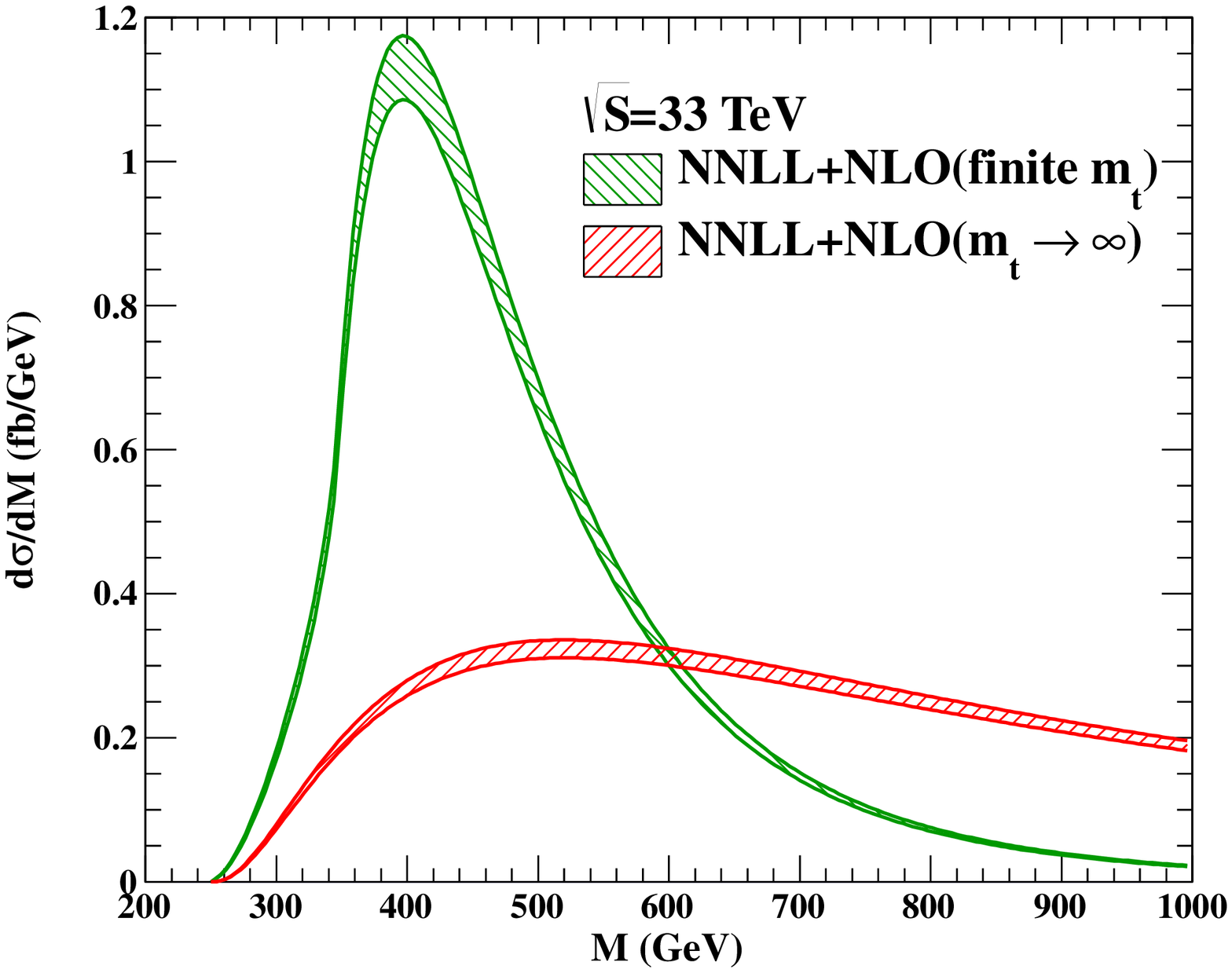}
\end{center}
\vspace{-4mm}
\caption{\label{invmass_reweight} RG-improved invariant mass distributions for Higgs boson pair production at the LHC with $\sqrt{S}=14$ and $33~{\rm TeV}$. The red bands are the differential distributions in the infinite top quark mass limit, and the green bands are differential distributions including finite top quark mass effects in the form factors. The black bands are the LO distributions including finite top quark mass by multiplied the K-factor obtained in the infinite top quark mass limits.
 }
\end{figure}

In Fig.~\ref{invmass_reweight} we compare the NNLL+NLO invariant mass distributions for Higgs boson pair production in the above two different approximations. In the first approximation we plot differential distributions in the infinite top quark mass limit~(red bands), and in the second approximation we plot differential distributions including exact top quark mass effects in the form factors~(green bands). It is shown that the red bands are extremely broad, while the green bands have peaks around about $M=400~$GeV. As mentioned before, in the first approximation we can not make correct predictions for the invariant mass distribution.

In Ref.~\cite{Grigo:2013rya} the total cross section of Higgs boson pair production as a function of the upper cut on the partonic center-of-mass energy are studied. Based on these results we can estimate that the top quark mass corrections may reach $\mathcal{O}(10\%)$ for the invariant mass distributions in the second approximation.

\subsection{Total cross section}
After performing integration over the Higgs boson pair invariant mass $M$, we can get the total cross sections of the Higgs boson pair invariant mass at the LHC. In Tab.~\ref{piresum_totxsec} we compare NLO singular cross section in the threshold region and the NNLL resummation results containing standard soft gluon resummation with $\mu_h^2=3.6M^2$, $\pi^2$-enhanced terms resummation with $\mu_h^2=-3.6M^2$ and the combination of both at the LHC with $\sqrt{S}=14$ and $33$~TeV, respectively. As is shown in Tab.~\ref{piresum_totxsec} that standard soft gluon resummation enhance the NLO singular cross section by about 10\%, and $\pi^2$-enhanced terms resummation enhance the NLO singular cross section by about 25\%. Thus in our RG-improved predictions the main contributions come from the resummation of the $\pi^2$-enhanced terms. Besides, similarly to the previous cases of the invariant mass distributions, the standard soft gluon resummation enhanced effects decrease and $\pi^2$-enhanced effects hardly change with the increasing of center-of-mass energy.

\begin{table}[h]
\vspace{0.2cm}
\centering
\begin{tabular}{ccccc}
\hline
$\sqrt{S}$ [TeV] & Singular [fb] & Standard [fb]& $\pi^2$-Enhanced [fb] & Both [fb]
\vspace{0.2cm}\\
\hline
\hline
\\[-0.4cm]
14 &
$30.7^{+3.4}_{-3.4}$ &
$34.3^{+3.6}_{-1.8}$ &
$38.7^{+3.2}_{-4.2}$ &
$41.4^{+3.2}_{-2.2}$
\vspace{0.2cm}\\
33 &
$182.9^{+13.2}_{-14.4}$ &
$197.2^{+22.2}_{-13.1}$ &
$228.5^{+17.2}_{-16.7}$ &
$238.3^{+20.9}_{-16.2}$
\vspace{0.2cm}\\
\hline
\end{tabular}
\caption{\label{piresum_totxsec}
Cross sections of Higgs boson pair production at the LHC. We present NLO cross section in the threshold region and the NNLL resummation results containing standard soft gluon resummation with $\mu_h^2=3.6M^2$, $\pi^2$-enhanced terms resummation with $\mu_h^2=-3.6M^2$ and the combination of both. The errors are the scale uncertainties.}
\end{table}

In Tab.~\ref{totxsec} we show the NLO and NNLL+NLO total cross section of Higgs boson pair production for different Higgs boson mass at the LHC with $\sqrt{S}=14~ {\rm and}~ 33$ TeV. The first errors of the cross sections represent the scale uncertainties associated with $\mu_h$, $\mu_s$ and $\mu_f$. In the NLO results the second errors are the PDF+$\as$ uncertainties from MSTW2008NLO 90\%~(68\%) Confidence Level (CL) PDF sets and in the NNLL+NLO results the second errors are the PDF+$\as$ uncertainties form MSTW2008NNLO 90\%~(68\%) CL PDF sets, where the PDF+$\as$ uncertainties are obtained using the formulas in Ref.~\cite{Martin:2009bu}. Our results show that the resummation effects increase the NLO results by about $27\%$ and $22\%$ with $\sqrt{S}=14~ {\rm and}~ 33$ TeV, respectively. At the NLO the uncertainties mainly come from scale uncertainties and are about $30\%$. The PDF+$\as$ uncertainties are about $12\%$ and $6\%$ for MSTW2008NNLO 90\% and 68\% CL, respectively. After performing the resummations the scale uncertainties are reduced to $8\%$ and the PDF+$\as$ uncertainties are almost unchanged in percentages. Another uncertainties coming from top quark mass corrections \cite{Grigo:2013rya} are about $\mathcal{O}(10\%)$ as mentioned above.

\begin{table}
\vspace{0.2cm}
\centering
\begin{tabular}{cccc}
\toprule[2pt]
\multirow{2}{*}{$m_{\rm H}$~[GeV]} & \multicolumn{3}{c}{$\sqrt{S}=14~{\rm TeV}$} \\[-0.1cm]
  & NLO~[fb] & NLO + NNLL~[fb] & K-factor \\
\hline
\\[-0.4cm]
121 &
 $35.6^{+6.4 +2.4\,(+1.1)}_{-5.3 -2.1\,(-0.9)}$ &
 $45.1^{+2.7+3.4\,(+1.7)}_{-0.8-3.3\,(-1.4)}$ &
 1.27
\vspace{0.2cm}\\
123 &
 $34.7^{+6.3 +2.4 \, (+1.0)}_{-5.2 - 2.1 \, (-0.9)}$ &
 $44.0^{+2.6 +3.4 \, (+1.6)}_{-0.8 - 3.2 \, (-1.4)}$ &
 1.27
\vspace{0.2cm}\\
125 &
 $33.9^{+6.1+2.3 \, (+1.0)}_{-5.0 - 2.0 \, (-0.9)}$ &
 $42.9^{+2.6+3.3 \, (+1.6)}_{-0.8 - 3.1 \, (-1.3)}$ &
 1.27
\vspace{0.2cm}\\
127 &
 $33.0^{+6.0 +2.3 \, (+1.0)}_{-4.9 - 2.0 \, (-0.9)}$ &
 $42.0^{+2.5+3.2 \, (+1.6)}_{-0.8 - 3.0 \, (-1.3)}$ &
 1.27
 \vspace{0.2cm}\\
129 &
 $32.2^{+5.8 +2.2 \, (+1.0)}_{-4.8 - 1.9 \, (-0.8)}$ &
 $40.9^{+2.4+3.1 \, (+1.5)}_{-0.8 - 3.0 \, (-1.3)}$ &
 1.27
\vspace{0.2cm}\\
\midrule[1.5pt]
\multirow{2}{*}{$m_{\rm H}$~[GeV]}  & \multicolumn{3}{c}{$\sqrt{S}=33~{\rm TeV}$} \\[-0.1cm]
  & NLO~[fb] & NLO + NNLL~[fb] & K-factor \\
\hline
\\[-0.4cm]
121 &
 $216.7^{+32.9+13.4\,(+5.6)}_{-27.1-12.2\,(-5.1)}$ &
 $263.5^{+16.0+16.2\,(+9.0)}_{-4.9-15.6\,(-6.4)}$ &
 1.22
\vspace{0.2cm}\\
123 &
 $211.9^{+32.2+13.1 \, (+5.6)}_{-26.5 - 11.9 \, (-5.0)}$ &
 $257.7^{+15.6+16.2 \, (+8.8)}_{-4.8 - 15.3 \, (-6.3)}$ &
 1.22
\vspace{0.2cm}\\
125 &
 $207.2^{+31.5+12.8 \, (+5.5)}_{-25.9 - 11.6 \, (-4.9)}$ &
 $252.0^{+15.2+15.8 \, (+8.6)}_{-4.7 - 14.9 \, (-6.1)}$ &
 1.22
\vspace{0.2cm}\\
127 &
 $202.5^{+30.8+12.5 \, (+5.4)}_{-25.4 - 11.4 \, (-4.7)}$ &
 $246.3^{+14.9+15.4 \, (+8.4)}_{-4.6 - 14.6 \, (-6.0)}$ &
 1.22
 \vspace{0.2cm}\\
129 &
 $197.9^{+30.1+12.2 \, (+5.2)}_{-24.8 - 11.1 \, (-4.6)}$ &
 $240.7^{+14.5+15.1 \, (+8.2)}_{-4.5 - 14.3 \, (-5.9)}$ &
 1.22
\vspace{0.2cm}\\
\bottomrule[2pt]
\end{tabular}
\caption{\label{totxsec}
NLO and NLO+NNLL total cross sections of Higgs boson pair production at the LHC for different Higgs boson mass. The first errors represent the scale uncertainties. In the NLO results the second errors are the PDF+$\as$ uncertainties form MSTW2008NLO 90\%~(68\%) CL PDF sets and in the RG-improved results the second errors are the PDF+$\as$ uncertainties form MSTW2008NNLO 90\%~(68\%) CL PDF sets. }
\end{table}

Once the full NLO QCD corrections of the Higgs boson production including exact top quark mass effects are available in the future, then the resummation calculations can be updated immediately. We expect that the standard soft gluon resummation effects based on full NLO QCD calculations will not be changed due to the fact that these resummation enhanced effects come from the soft gluon radiation in the initial state and are independent of infinite top quark mass limit. If the exact top quark mass effects were included, the analytical expressions of hard functions would be changed. However, the squared logarithmic terms in Eq.~(\ref{hardcs}) will remain unchanged. This is because that the scale dependent logarithmic terms associated with infrared divergences, and the infrared behaviors of this process are independent of infinite top quark mass limit. Hence, the resummation effects of $\pi^2$-enhanced terms will also not be changed.

\begin{table}
\vspace{0.2cm}
\centering
\begin{tabular}{cccc}
\toprule[2pt]
\multirow{2}{*}{$\lambda/\lambda_{\rm SM}$} & \multicolumn{3}{c}{$\sqrt{S}=14~{\rm TeV}$} \\[-0.1cm]
  & NLO~[fb] & NLO + NNLL~[fb] & K-factor \\
\hline
\\[-0.4cm]
-1 &
 $127.9^{+23.1+8.7\,(+3.8)}_{-18.8-7.7\,(-3.3)}$ &
 $161.6^{+9.8+12.0\,(+6.0)}_{-3.1-11.4\,(-4.9)}$ &
 1.26
\vspace{0.2cm}\\
-0.8 &
 $115.0^{+20.7+7.8\,(+3.4)}_{-17.0-6.9\,(-2.9)}$ &
 $145.3^{+8.8+10.8\,(+5.3)}_{-2.8-10.2\,(-4.4)}$ &
 1.26
\vspace{0.2cm}\\
-0.6 &
 $102.8^{+18.6+7.0\,(+3.0)}_{-15.2-6.2\,(-2.6)}$ &
 $130.0^{+7.8+9.7\,(+4.8)}_{-2.5-9.2\,(-3.9)}$ &
 1.26
\vspace{0.2cm}\\
-0.4 &
 $91.5^{+16.5+6.2\,(+2.7)}_{-13.5-5.5\,(-2.3)}$ &
 $115.7^{+7.0+8.6\,(+4.3)}_{-2.2-8.2\,(-3.5)}$ &
 1.26
\vspace{0.2cm}\\
-0.2 &
 $80.9^{+14.6+5.5\,(+2.4)}_{-12.0-4.8\,(-2.1)}$ &
 $102.4^{+6.2+7.7\,(+3.8)}_{-1.9-7.3\,(-3.1)}$ &
 1.26
\vspace{0.2cm}\\
0 &
 $71.1^{+12.8+4.8 \, (+2.1)}_{-10.5 - 4.3 \, (-1.8)}$ &
 $90.0^{+5.4+6.8 \, (+3.3)}_{-1.7 - 6.4 \, (-2.8)}$ &
 1.27
\vspace{0.2cm}\\
0.2 &
 $62.1^{+11.2+4.2\,(+1.8)}_{-9.2-3.7\,(-1.6)}$ &
 $78.6^{+4.7+5.9\,(+2.9)}_{-1.5-5.6\,(-2.4)}$ &
 1.27
\vspace{0.2cm}\\
0.4 &
 $53.9^{+9.7+3.7\,(+1.6)}_{-8.0-3.2\,(-1.4)}$ &
 $68.2^{+4.1+5.2\,(+2.5)}_{-1.3-4.9\,(-2.1)}$ &
 1.27
\vspace{0.2cm}\\
0.6 &
 $46.4^{+8.4+3.2\,(+1.4)}_{-6.9-2.8\,(-1.2)}$ &
 $58.8^{+3.5+4.5\,(+2.2)}_{-1.1-4.2\,(-1.8)}$ &
 1.27
\vspace{0.2cm}\\
0.8 &
 $39.8^{+7.2+2.7\,(+1.2)}_{-5.9-2.4\,(-1.0)}$ &
 $50.4^{+3.0+3.8\,(+1.9)}_{-0.9-3.6\,(-1.6)}$ &
 1.27
\vspace{0.2cm}\\
1 &
 $33.9^{+6.1 + 2.3 \, (+1.0)}_{-5.0 - 2.0 \, (-0.9)}$ &
 $42.9^{+2.6 + 3.3 \, (+1.6)}_{-0.8 - 3.1 \, (-1.3)}$ &
 1.27
\vspace{0.2cm}\\
1.2 &
 $28.8^{+5.2+2.0\,(+0.9)}_{-4.3-1.7\,(-0.7)}$ &
 $36.5^{+2.2+2.8\,(+1.4)}_{-0.7-2.7\,(-1.2)}$ &
 1.27
\vspace{0.2cm}\\
1.4 &
 $24.4^{+4.4+1.7\,(+0.7)}_{-3.6-1.5\,(-0.6)}$ &
 $31.0^{+1.9+2.4\,(+1.2)}_{-0.6-2.3\,(-1.0)}$ &
 1.27
\vspace{0.2cm}\\
1.6 &
 $20.9^{+3.8+1.4\,(+0.6)}_{-3.1-1.3\,(-0.5)}$ &
 $26.5^{+1.6+2.1\,(+1.0)}_{-0.5-1.9\,(-0.8)}$ &
 1.27
\vspace{0.2cm}\\
1.8 &
 $18.1^{+3.3+1.2\,(+0.5)}_{-2.7-1.1\,(-0.5)}$ &
 $23.0^{+1.4+1.8\,(+0.9)}_{-0.4-1.7\,(-0.7)}$ &
 1.27
\vspace{0.2cm}\\
2 &
 $16.1^{+2.9+1.1 \, (+0.5)}_{-2.4 - 1.0 \, (-0.4)}$ &
 $20.4^{+1.2+1.6 \, (+0.8)}_{-0.4 - 1.5 \, (-0.7)}$ &
 1.27
\vspace{0.2cm}\\
\bottomrule[2pt]
\end{tabular}
\caption{\label{lam_totxsec_1}
NLO and RG-improved total cross sections of Higgs boson pair production at the LHC with $\sqrt{S}=14$~TeV for different Higgs boson self-coupling $\lambda$. The first errors represent the scale uncertainties. The second errors are PDF+$\as$ uncertainties. }
\end{table}

\begin{table}
\vspace{0.2cm}
\centering
\begin{tabular}{cccc}
\toprule[2pt]
\multirow{2}{*}{$\lambda/\lambda_{\rm SM}$} & \multicolumn{3}{c}{$\sqrt{S}=33~{\rm TeV}$} \\[-0.1cm]
  & NLO~[fb] & NLO + NNLL~[fb] & K-factor \\
\hline
\\[-0.4cm]
-1 &
 $725.6^{+109.8+45.5\,(+19.4)}_{-89.7-41.7\,(-17.4)}$ &
 $881.4^{+54.2+55.4\,(+30.8)}_{-16.5-52.4\,(-21.3)}$ &
 1.21
\vspace{0.2cm}\\
-0.8 &
 $655.3^{+99.1+41.0\,(+17.4)}_{-81.1-37.6\,(-15.8)}$ &
 $796.0^{+48.9+50.0\,(+27.8)}_{-14.9-47.3\,(-19.3)}$ &
 1.21
\vspace{0.2cm}\\
-0.6 &
 $589.0^{+89.1+36.9\,(+15.7)}_{-72.9-33.7\,(-14.1)}$ &
 $715.6^{+43.9+44.9\,(+24.9)}_{-13.4-42.5\,(-17.3)}$ &
 1.21
\vspace{0.2cm}\\
-0.4 &
 $526.9^{+79.8+32.9\,(+14.0)}_{-65.2-30.1\,(-12.6)}$ &
 $640.2^{+39.2+40.2\,(+22.3)}_{-12.0-38.0\,(-15.5)}$ &
 1.22
\vspace{0.2cm}\\
-0.2 &
 $468.8^{+71.0+29.3\,(+12.5)}_{-58.1-26.8\,(-11.2)}$ &
 $569.7^{+34.9+35.8\,(+19.8)}_{-10.6-33.8\,(-13.8)}$ &
 1.22
\vspace{0.2cm}\\
0 &
 $414.9^{+62.9+25.9 \, (+11.0)}_{-51.5 - 23.6 \, (-9.9)}$ &
 $504.3^{+30.8+31.6 \, (+17.5)}_{-9.4 - 30.0 \, (-12.2)}$ &
 1.22
\vspace{0.2cm}\\
0.2 &
 $365.2^{+55.4+22.7\,(+9.7)}_{-45.4-20.8\,(-8.7)}$ &
 $443.8^{+27.1+27.9\,(+15.4)}_{-8.3-26.3\,(-10.8)}$ &
 1.22
\vspace{0.2cm}\\
0.4 &
 $319.5^{+48.5+19.8\,(+8.5)}_{-39.8-18.1\,(-7.6)}$ &
 $388.4^{+23.7+24.4\,(+13.4)}_{-7.2-23.0\,(-9.4)}$ &
 1.22
\vspace{0.2cm}\\
0.6 &
 $277.9^{+42.2+17.2\,(+7.4)}_{-34.7-15.7\,(-6.6)}$ &
 $337.9^{+20.5+21.2\,(+11.6)}_{-6.3-20.0\,(-8.2)}$ &
 1.22
\vspace{0.2cm}\\
0.8 &
 $240.5^{+36.5+14.9\,(+6.4)}_{-30.0-13.5\,(-5.7)}$ &
 $292.4^{+17.7+18.3\,(+10.1)}_{-5.4-17.3\,(-7.1)}$ &
 1.22
\vspace{0.2cm}\\
1 &
 $207.2^{+31.5+12.8 \, (+5.5)}_{-25.9 - 11.6 \, (-4.9)}$ &
 $252.0^{+15.2+15.8 \, (+8.6)}_{-4.7 - 14.9 \, (-6.1)}$ &
 1.22
\vspace{0.2cm}\\
1.2 &
 $178.0^{+27.1+11.0\,(+4.7)}_{-22.3-10.0\,(-4.2)}$ &
 $216.5^{+13.1+13.6\,(+7.4)}_{-4.0-12.8\,(-5.3)}$ &
 1.22
\vspace{0.2cm}\\
1.4 &
 $152.9^{+23.3+9.4\,(+4.0)}_{-19.2-8.5\,(-3.6)}$ &
 $186.0^{+11.2+11.7\,(+6.3)}_{-3.4-11.0\,(-4.6)}$ &
 1.22
\vspace{0.2cm}\\
1.6 &
 $131.9^{+20.1+8.1\,(+3.5)}_{-16.6-7.3\,(-3.1)}$ &
 $160.5^{+9.6+10.1\,(+5.5)}_{-3.0-9.5\,(-3.9)}$ &
 1.21
\vspace{0.2cm}\\
1.8 &
 $115.0^{+17.6+7.1\,(+3.0)}_{-14.5-6.4\,(-2.7)}$ &
 $139.9^{+8.4+8.8\,(+4.8)}_{-2.6-8.3\,(-3.4)}$ &
 1.22
\vspace{0.2cm}\\
2 &
 $102.3^{+15.7+6.3 \, (+2.7)}_{-12.9 - 5.7 \, (-2.4)}$ &
 $124.4^{+7.4+7.8 \, (+4.2)}_{-2.3 - 7.4 \, (-3.1)}$ &
 1.22
\vspace{0.2cm}\\
\bottomrule[2pt]
\end{tabular}
\caption{\label{lam_totxsec_2}
NLO and RG-improved total cross sections of Higgs boson pair production at the LHC with $\sqrt{S}=33$~TeV for different Higgs boson self-coupling $\lambda$. The first errors represent the scale uncertainties. The second errors are PDF+$\as$ uncertainties. }
\end{table}

\subsection{Higgs boson self-coupling measurements at the LHC}

Once the discovery of the Higgs-like particle is confirmed, it is important to verify whether it is the SM Higgs or not. Thus it is necessary to measure the Higgs boson self-coupling $\lambda$ at the LHC through Higgs boson pair production. In the expression of effective Lagrangian (\ref{effL}) the first and the second operator induce the first and second Feynman diagram in Fig.~\ref{born}, respectively, where only the physical process represented by the first Feynman diagrams is sensitive to the value of Higgs boson self-coupling. Due to the different sign of these two terms in Eq.~(\ref{effL}) important interference effects exist. If $\lambda<0$, the contributions from these two terms can be added, and if $\lambda>0$, the contributions cancel each other. Thus the total cross section of Higgs boson pair production decreases as increasing of $\lambda$. This relation is shown in Tab.~\ref{lam_totxsec_1} and Tab.~\ref{lam_totxsec_2}, where we present the dependence of the Higgs boson pair production cross section on the self-coupling $\lambda$. The scale and PDF+$\as$ uncertainties are both listed. Obviously, the theoretical predictions at the NLO level are not sufficient to precisely determine the value of $\lambda$ due to high scale dependence and underestimate of the total cross section. After including the NNLL resummation effects, it is promising to distinguish the SM Higgs boson self-couplings from non-SM ones.
\begin{figure}
\begin{center}
\includegraphics[width=1\textwidth]{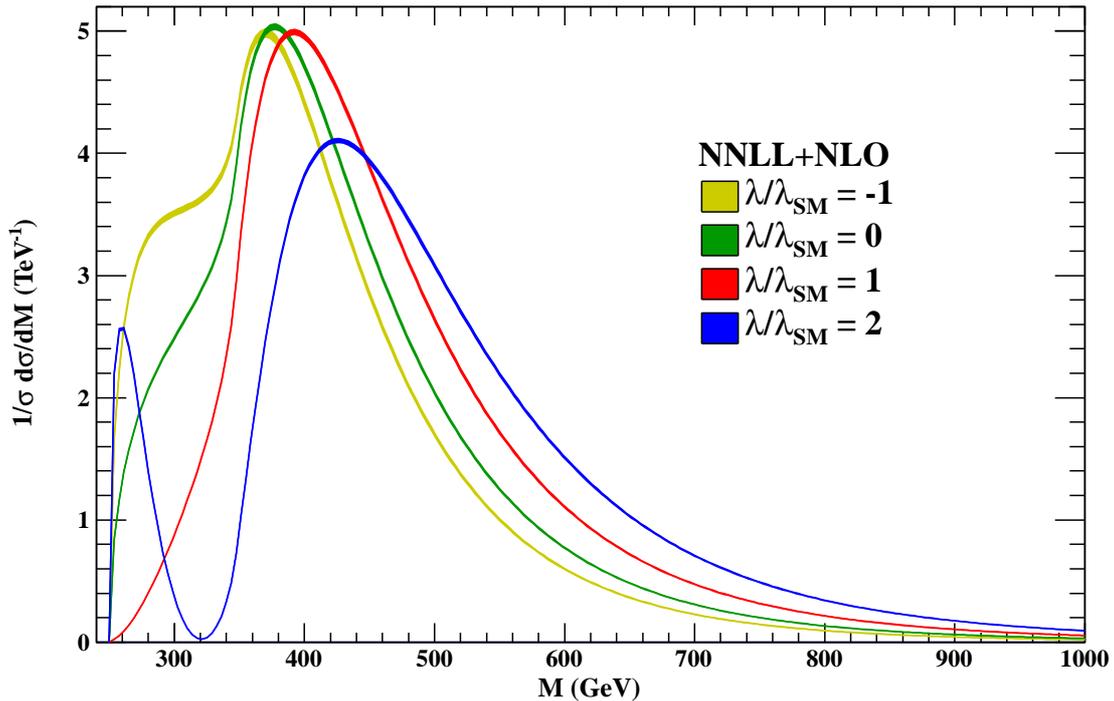}
\end{center}
\vspace{-4mm}
\caption{\label{lam_mHH}
The normalized Higgs boson pair invariant mass distribution at the LHC with $\sqrt{S}=14$ TeV, where the bands represent the scale uncertainties.}
\end{figure}

Moreover, these important interference effects also contribute to the invariant mass distribution of the Higgs boson pair production. In Fig.~\ref{lam_mHH}, we present the normalized Higgs boson pair invariant mass distribution at the LHC with $\sqrt{S}=14$ TeV where the scale uncertainties almost vanish due to the cancelation of scale uncertainties in ratio. When $\lambda<0$, there exists enhancement effects in the invariant mass region of $300~$GeV~(yellow band), compared to the cases of $\lambda=0$~(green band) and $\lambda_{\rm SM}$~(red band). With the increasing of the value of $\lambda$, these enhancement effects decrease and the position of peak moves towards large invariant mass region. Especially, when $\lambda>\lambda_{\rm SM}$ there exits two different peak~(blue band). In order to gain some insight into the reason why the shape of the invariant mass distribution strongly depends on the value of $\lambda$, it is helpful to consider the limit of the infinite top quark mass. As is shown in Eq.(\ref{cs}) the differential cross sections in the infinite top quark mass limit share a common factor
\begin{eqnarray}
 1 - \frac{6\lambda v^2}{M^2 - m_H^2 + i m_H \Gamma_H},
\end{eqnarray}
where the second terms represent a Higgs boson exchanged in the s-channel. When $\lambda<0$, the invariant mass distributions at the small value region are added up, thus an obvious enhancement effects exists in the small invariant mass region. When $\lambda>0$, in the small invariant mass region the contributions from two terms cancel each other, so these enhancement effects decrease and the peak of the invariant mass distribution moves toward large value region with the increasing value of $\lambda$. In particular, when $M^2 = 6\lambda v^2+m_H^2$ the common factor equals zero. Considering the threshold limit $M>2m_H$, there exists a zero point in the invariant mass distribution as $\lambda>\lambda_{\rm SM}$. For example, as shown in Fig.~\ref{lam_mHH}, there are two peaks in the invariant mass distribution when $\lambda=2\lambda_{\rm SM}$, where the differential distribution equals zero with $M=330.7$ GeV.

Our above discussions about the dependence of Higgs boson self coupling on the total cross section and invariant mass distribution are based on the approximated method of using the form factor to contain parts of top quark mass effects. Therefore our theoretical predictions receive about $\mathcal{O}(10\%)$ uncertainties~\cite{Grigo:2013rya}. However once the full NLO QCD corrections of the Higgs boson production including exact top quark mass effects are available in the future, the dependence of the resummed total cross section and invariant mass distribution on the Higgs boson self coupling can be updated immediately, and we can make more precise predictions. On the other side, above discussions provide some important information about the properties of the Higgs boson pair invariant mass distribution shape. Especially, we see that it is possible to extract the parameter $\lambda$ from the total cross section and Higgs boson pair invariant mass distribution when the measurement precision increases at the LHC.

\section{CONCLUSION}\label{s5}
We have calculated the resummation effects in the SM Higgs boson pair production at the LHC with SCET. We present the invariant mass distribution and the total cross section at NNLL level with $\pi^2$-enhanced terms resummed, which are matched to the NLO results. In the high order QCD predictions full form factors including exact top quark mass effects are used. Our results show that the resummation effects increase the NLO results by about $20\% \sim 30\%$, and the scale uncertainty is reduced to $8\%$, which leads to increased confidence on the theoretical predictions. The theoretical uncertainties from the top quark mass effects are about $\mathcal{O}(10\%)$. The PDF+$\as$ uncertainties are almost not changed after including resummation effects. We also study the sensitivity of the total cross section and the invariant mass distribution to the Higgs boson self-coupling. We find that the NNLL resummation effects can decrease the needed luminosity to detect the Higgs boson self-coupling due to the enhancement effects of the total cross section, and the total cross section and the invariant mass distribution shape depend strongly on the self-coupling. As a result, it is possible to extract Higgs boson self-coupling from the total cross section and Higgs boson pair invariant mass distribution when the measurement precision increases at the LHC.

\begin{acknowledgments}
We would like to thank Qing-Hong Cao and Li Lin Yang for helpful discussions. This work was supported in part by the National Natural Science Foundation of China under Grants No. 11021092 and No. 11135003.
\end{acknowledgments}

\appendix
%%%%%%%%%%%%%%%%%%%%%%%%%%%% appendix A %%%%%%%%%%%%%%%%%%%%%%%
\section{THE EXPRESSIONS OF THE FORM FACTORS}\label{a1}
The definitions of the triangle and box form factors $f_{\rm Tri}^{\rm A}$, $f_{\rm Box}^{\rm A}$ and $f_{\rm Box}^{\rm B}$ in Eq.~\ref{fin_mt_born} can be written as
\begin{eqnarray}\label{formfac1}
 f_{\rm Tri}^{\rm A}(s,t,u,m_t^2,m_H^2) &=& \frac{18\lambda m_t^2 v^2}{s(s - m_H^2 + i m_H \Gamma_H)}\left[ 2 + (4m_t^2 - s) C_0^s \right],  \\
 f_{\rm Box}^{\rm A}(s,t,u,m_t^2,m_H^2) &=& \frac{3 m_t^2}{s}\Big\{ m_t^2(8m_t^2 - s -2m_H^2 )(D_0^u + D_0^t + D_0^{tu}) + p_T^2(4m_t^2 - m_H^2) D_0^{tu} \nno \\
 && + 2 + 4m_t^2 C_0^s + \frac{2}{s}(m_H^2 - 4 m_t^2)\left[ (t - m_H^2) C_0^t + (u - m_H^2) C_0^u \right] \Big\}, \nno \\
 \\
 f_{\rm Box}^{\rm B}(s,t,u,m_t^2,m_H^2) &=& \frac{3 m_t^2}{2 s}\Big\{ 2(8 m_t^2 + s - 2 m_H^2)\left[ m_t^2(D_0^u + D_0^t + D_0^{tu}) - C_0^{\rm sm} \right] \nno \\
 && - 2\left[ s C_0^s + (t - m_H^2)C_0^t + (u - m_H^2)C_0^u \right] \nno \\
 && + \frac{1}{s p_T^2}\Big[ s u (8 u m_t^2 - u^2 - m_H^4) D_0^u + s t (8 t m_t^2 - t^2 - m_H^4) D_0^t \nno \\
 && + (8m_t^2 + s - 2m_H^2)[ s(s-2 m_H^2) C_0^s + s(s-4m_H^2)C_0^{sm} \nno \\
 && + 2 t(m_H^2 - t) C_0^t + 2 u (m_H^2 - u)C_0^u ] \Big] \Big\},
\end{eqnarray}
where we have defined some convenient symbols to represent the corresponding scalar functions as follows,
\begin{align}
 C_0^s &= C_0(0,0,s;m_t^2,m_t^2,m_t^2), ~~~ C_0^{sm} = C_0(m_H^2, s, m_H^2, m_t^2, m_t^2, m_t^2), \nno \\
 C_0^t &= C_0(0, t, m_H^2; m_t^2, m_t^2, m_t^2), ~~~ C_0^u = C_0(0, u, m_H^2; m_t^2, m_t^2, m_t^2), \nno \\
 D_0^t &= D_0(m_H^2, 0, 0, m_H^2; t, s; m_t^2, m_t^2, m_t^2, m_t^2), \nno \\
 D_0^u &= D_0(m_H^2, 0, 0, m_H^2; u, s; m_t^2, m_t^2, m_t^2, m_t^2), \nno \\
 D_0^{tu} &= D_0(m_H^2, 0, m_H^2, 0; t, u; m_t^2, m_t^2, m_t^2, m_t^2).
\end{align}
The definitions of the form factors $g^{\rm A}$ and $g^{\rm B}$ in Eq.~\ref{cij_def} are
\begin{align}
 g^{\rm A} &= \frac{3}{64} \left( C_t^2 + C_u^2 \right), \nno \\
 g^{\rm B} &= \frac{3 p_T^2}{64} \left( \frac{C_t^2}{t} + \frac{C_u^2}{u} \right),
\end{align}
where
\begin{align}
 C_t &= \frac{16 m_t^2 t}{(m_H^2 -t)^2} \left[ B_0(m_H^2, m_t^2, m_t^2) - B_0(t, m_t^2, m_t^2) \right] + \frac{16 m_t^2}{(m_H^2 - t)} \nno \\
 & + 4 m_t^2 \left( \frac{8 m_t^2}{m_H^2 - t} - 2 \right) C_0(0,t,m_H^2;m_t^2,m_t^2,m_t^2), \\
 C_u &= C_t(t \leftrightarrow u).
\end{align}
The definitions of the scalar functions are listed as follows,
\begin{align}
 & B_0(p_1^2, m_t^2, m_t^2) = \frac{1}{i \pi^2} \int d^4 l \frac{1}{\left[l^2 - m_t^2\right] \left[(l + q_1)^2 - m_t^2\right]}, \nno \\
 & C_0(p_1^2, p_2^2, p_3^2; m_t^2, m_t^2, m_t^2) = \frac{1}{i \pi^2} \int d^4 l \frac{1}{\left[l^2 - m_t^2\right] \left[(l + q_1)^2 - m_t^2\right] \left[ (l + q_2)^2 - m_t^2 \right] }, \nno \\
 &D_0(p_1^2, p_2^2, p_3^2, p_4^2; s_{12}, s_{23}; m_t^2, m_t^2, m_t^2, m_t^2) \nno \\
 &= \frac{1}{i \pi^2} \int d^4 l \frac{1}{ \left[l^2 - m_t^2\right] \left[(l + q_1)^2 - m_t^2\right] \left[ (l + q_2)^2 - m_t^2 \right] \left[ (l + q_3)^2 - m_t^2 \right] },
\end{align}
where $q_n = \sum_{i=1}^n p_i$ and $s_{ij} = (p_i + p_j)^2$. In our numerical calculations, these loop integrations are evaluated with ONELOOP~\cite{vanHameren:2010cp}. In the infinite top quark mass limit, one can make an expansion in inverse powers of the top quark mass. Thus, the scalar functions are given by
\begin{align}
&B_0(p_1^2, m_t^2, m_t^2) = \frac{1}{\epsilon} - \gamma_{\rm E} - \ln\pi - \ln m_t^2 + \frac{p_1^2}{6 m_t^2} + \mathcal{O}(\frac{1}{m_t^4}) ,  \\
&C_0(p_1^2, p_2^2, p_3^2; m_t^2, m_t^2, m_t^2) = - \frac{1}{2 m_t^2} + \frac{ - p_1^2 - p_2^2 - p_3^2 }{24 m_t^4} + \mathcal{O}(\frac{1}{m_t^6}),  \\
&D_0(p_1^2, p_2^2, p_3^2, p_4^2; s_{12}, s_{23}; m_t^2, m_t^2, m_t^2, m_t^2) = \frac{1}{6 m_t^4} + \frac{p_1^2 + p_2^2 + p_3^2 + p_4^2 + s_{12} + s_{23}}{60 m_t^6} + \mathcal{O}(\frac{1}{m_t^8}),
\end{align}
where $\gamma_{\rm E}$ is the Euler constant.

\end{document}